\RequirePackage{snapshot}

\documentclass[12pt,epsfig]{article}
\usepackage[T1]{fontenc}
\usepackage[utf8]{inputenc}

\usepackage[dvips]{graphicx}
\usepackage{amsmath,amssymb}

\begin{document}

\begin{titlepage}
\title{Cosmological model with Born-Infeld type scalar field}

\author{{ A. Troisi$^{\diamond\,\star}$, \,E. S\'eri\'e$^{\star\,\dagger}$, \,\, R. Kerner$^{\star}$. \, }
\\
 \small $^{\diamond}$ Dipartimento di Scienze Fisiche, Universit\`a di
Napoli ``Federico II'',
 \\ \small
Universit\`{a} Compl. Univ. di Monte S. Angelo, Edificio G, Via Cinthia,
80121 - Napoli, Italy
\\
\small
 INFN, Sezione di Napoli, Compl. Univ. di Monte S. Angelo, \\
\small Edificio G, Via Cinthia, 80121 - Napoli, Italy
\\
 \small $^{\star}$
  Laboratoire de Physique Th\'eorique de la Mati\`ere Condens\'ee, \\\small
  Universit\'e Pierre-et-Marie-Curie - CNRS UMR 7600 \\\small
  Tour 22, 4-\`eme \'etage, Bo\^{i}te 142, \\\small
  4, Place Jussieu, 75005 Paris, France\\\small
  $^{\dagger}$
  Laboratoire de Physique Th\'eorique (UMR 8627)\\\small
  Universit\'e Paris XI,\\\small
  B\^atiment 210, 91405 Orsay Cedex, France
}

\maketitle


\begin{abstract}
The non-abelian generalization of the Born-Infeld non-linear lagrangian 
is extended to the non-commutative geometry of matrices on a manifold. 
In this case not only the usual $SU(n)$ gauge fields appear, but also a natural 
generalization of the multiplet of scalar 
Higgs fields, with the double-well potential as a first approximation.
 		 
The matrix realization of non-commutative geometry provides a natural framework 
in which the notion of a determinant can be easily generalized 
and used as the lowest-order term in a gravitational lagrangian of a new kind. 
As a result, we obtain a Born-Infeld-like lagrangian as a root of sufficiently 
high order of a combination of metric, gauge potentials and the scalar field interactions.

We then analyze the behavior of cosmological models based on this lagrangian.
It leads to primordial inflation with varying speed, with possibility of early deceleration 
ruled by the relative strength of the Higgs field.

\end{abstract}

\end{titlepage}

\section{Introduction}

\subsection{The origin of the Born-Infeld lagrangian}

\indent
Among numerous cosmological models using scalar field as a source
of primordial cosmic energy and the subsequent inflation there is
usually no limit on the field strength: as a matter of fact, the
scalar field $\Phi$ can take on arbitrarily high values. This is
true in particular for the simplest model with $V(\phi)=\lambda\phi^2\left(\phi^2\,-\,\gamma^2\right)$ ``double well'' potential.
However, one may point out that principles of
quantum field theory, when combined with General Relativity,
should lead to the existence of a cut-off for any field's value,
because at certain intensity the very notion of space and time
will not be valid anymore. More precisely, it is because Heisenberg's 
uncertainty principle induces a spontaneous birth of black holes, when applied to the gravitational field.

This is why it seems reasonable to investigate theories in which
such a cut-off is incorporated from the very beginning.
First such attempt concerned exclusively the electric field, whose infinite
value at $r\rightarrow 0$ in the Coulomb law should be avoided in
order to keep the energy finite.

To this purpose, G. Mie \cite{mie:12} introduced the notion of  {\it maximal field
strength}, ${\bf E}_0$,  in order to make it impossible for any
electric field to go beyond this value. He modified Maxwell's
theory by introducing the following non-linear lagrangian density
for pure electric field:
\begin{equation}
{\cal{L}} = \sqrt{1 - \frac{{\bf E}^2}{{\bf E}_0^2}}\,.
\label{Mie1}
\end{equation}
Although the non-linear theory derived from this lagrangian
enabled Mie to obtain a non-singular, finite energy solution, it
was clear that such a lagrangian can not represent a
Lorentz-invariant theory, especially that from the beginning there was
no magnetic field contribution at all. This is why Born and Infeld
\cite{born_infeld:34} introduced a Lorentz-invariant lagrangian
density (when constructed on a pseudo-Riemannian manifold, it is also
invariant under diffeomorphisms), defined as follows:
$$   {\cal{L}}_{BI}(g,F) = {L_{BI}(g,F)}
 \sqrt{|g|}= \beta^2 \Biggl( \sqrt{|\det(g_{\mu \nu}) |} -
  \sqrt{|\det(g_{\mu \nu}+\beta ^{-1} \, F_{\mu \nu}\,)\,|} \Biggr)  $$
\begin{equation}
 =  \beta^2\Biggl(1-\sqrt{1+ \frac{1}{\beta^{2}}
    ({\bf B}^2 - {\bf E}^2) - \frac{1}{\beta^{4}}({\bf E}\cdot {\bf B})^2}
\Biggr)
  \sqrt{|g|}\,.
\label{BI1}
\end{equation}
\noindent 
The constant $\beta$ appears for dimensional reasons
and plays the role of the upper limit of the electric
field in Mie's non-linear electrodynamics. Defining
$$  P = \frac{1}{4} \, F_{\mu \nu} \,
F^{\mu \nu}  \, \ \ {\rm and} \, \ \ \,  S = \frac{1}{4} \, F_{\mu
\nu} \, {\tilde{F}}^{\mu \nu} \, , \ \ {\rm with} \, \ \ \,
{\tilde{F}}^{\mu \nu} = \frac{1}{2} \, \, \epsilon^{\mu \nu
\lambda \rho } \, F_{\lambda \rho} $$ the Born-Infeld lagrangian
can be recast in a simple form:
\begin{equation}
{\cal L}_{BI}  = \beta^2 \, \biggl[ 1 - \sqrt{1 + 2 P - S^2}
\biggr] \label{kern-eq.1.3}
\end{equation}
\indent 
The idea of a non-abelian generalization of Born-Infeld
theory lagrangian has been in the air already at the end of the
seventies. Hagiwara has discussed various possibilities in
\cite{hagiwara:81}, however, he did not  try to find soliton-like
solutions. In 1997 Tseytlin \cite{tseytlin:97}  argued in favour
of the symmetrized trace prescription which reproduced in the
first 4 orders the string effective action for gauge potential.
Finally Park \cite{park:99} introduced yet another non-abelian
generalization and investigated qualitative  behavior of
instanton-like solutions.

It is interesting to note that the non-commutative generalizations
of geometry of the space-time lead quite naturally to this class of lagrangians.

\subsection{Arguments in favour of non-commutative geometry}
\indent
Simultaneous consideration of the two most important new physical theories
of this century, General Relativity and Quantum Mechanics, have not yet
produced a common tool for the description of the nature of space-time at the
microscopic level. The General Relativity develops our knowledge about global
properties of space and time at very large distances, and raises
questions concerning the global topology of the Universe. 
\newline
\indent
The methods of Differential Geometry, the best adapted 
mathematical language of this theory, are very different from the 
methods of Quantum Physics, in which one studies the properties of the 
algebra of observables, considering the state vectors, as well as geometric 
points and trajectories, as artefacts and secondary notions. This approach 
has been inspired by the works of John von Neumann [9], and has much in 
common with the non-commutative geometry, where the very notion of a point 
loses its meaning.

\indent
There are several well-known arguments in favour of the point of view
according to which the dynamical 
interplay between Quantum Theory and Gravitation should lead to a 
non-commutative version of space-time. Let us recall the most frequently
cited ones:
\vskip 0.2cm
\indent
\hskip 0.5cm
* A semi-classical argument that involves the black hole creation at very small
distances: as a matter of fact, if General Relativity remains valid at
the Planck scale, then any localization of events should become impossible
at the distances of the order of ${\lambda}_P = \sqrt{\frac{\hbar G}{c^3}}$.
Indeed, according to quantum mechanical principles, to localize an event
in space-time within the radius $\Delta \, x^{\mu} \sim a$, one needs to
employ energies of the order of $a^{-1}$. When $a$ becomes very small, the
creation of mini black holes becomes possible, thus excluding from the
observation that portion of the space-time and making further localization
meaningless.
Therefore, the localization is possible only if we impose the following 
limitation on the time interval:
\begin{equation}
\Delta \, x^0 \, ( \Sigma \, \Delta \, x^k) \, \geq {\lambda}^2_P \, \ \ \, 
{\rm and} \, \ \ \, \ \ \Delta \, x^k \, \Delta \, x^m \geq {\lambda}^2_P .
\end{equation}
in order to avoid the black hole creation at the microscopic level.
\vskip 0.2cm
\indent
\hskip 0.5cm
** The topology of the space-time should be sensitive to the states of the 
fields which are in presence - and {\it vice versa}, quantum evolution of
any field, including gravity, should take into account all possible field
configurations, also corresponding to the fields existing in space-times
with radically different topologies (a creation of a black hole is but the
simplest example; one should also take into account other ``exotic''
configurations, such as multiple Einstein-Rosen bridges (the so-called
``{\it wormholes}''), leading in the limit of great $N$ to the space-time 
{\it foam}). 
\newline
\indent
Now, as any quantum measurement process may also lead to topological 
modifications, again the coordinates of an event found before and after any
measurement can not be compared anymore because they might refer to uncompatible
coordinate patches in different local maps. As a result, quantum measures
of coordinates themselves become non-commutative, and the algebra of functions
on the space-time, supposed to contain also all possible local coordinates,
must be replaced by its non-commutative extension, better adapted to describe
the space-time foam.
\vskip 0.2cm
\indent
\hskip 0.5cm
*** Since the coordinates $x^{\mu}$ are endowed with a length scale, the
metric must enter at certain stage in order to measure it. After quantization,
the components of the tensor $g_{\mu \nu}$ become a set of dynamical fields, 
whose behavior is determined by the propagators and, at least at the lowest
perturbative level, by two-point correlation functions. As any other field,
the components of the metric tensor will display quantum fluctuations, making
impossible precise measurements of distances, and therefore, any precise
definition of coordinates. Let us now consider possible consequences of the above 
assumptions.

\section{The Non-commutative Born-Infeld lagrangian}

\subsection{Non-Commutative Einstein-Hilbert lagrangian}

Let us assume therefore that at the Planck scale not only the positions and momenta 
do not commute anymore, but also the coordinates themselves should belong to 
a non-commutative algebra. As a direct consequence, all functions of coordinates 
that served to describe various geometrical quantities should also belong to this algebra, 
although some of them can belong to its center and commute between themselves. 

The non-commutativity of coordinates means also that the notion of distance between
two points of space-time can not remain symmetric: if the quantum indetermination
principle applies to the measurements of positions, then there is no guarantee that
the distance mesured first from a point $A$ to its neighbor $B$ will remain strictly
the same when we try to measure next the distance from $B$ to $A$. This amounts to
the introduction of a non-symmetric metric tensor, $g_{ij} \neq g_{ji}$. In his efforts
to create a general field theory unifying gravity with other fields Einstein considered
such a possibility in his late years, but without quantum theoretical motivation,
and his investigations remained purely classical. 

Now, combining two direct consequences of non-commutative character of space-time coordinates
and their functions leads quite naturally to the following general form of the metric tensor:
\begin{equation}
g_{ij} = g^0_{ij} + {\hat{G}_{ij}} + f^0_{ij} + {\hat{F}_{ij}},
\label{gijgeneral}
\end{equation}
In this formula we have separated the commutative and non-commutative parts, as well as
the symmetric and antisymmetric parts: the terms $g^0_{ij}$ and $f^0_{ij}$ are supposed
to belong to the center of our hypothetic non-commutative algebra replacing the usual
algebra of smooth functions of coordinates, with $g^0_{ij} = g^0_{ij}$ and $f^0_{ij} = - f^0_{ji}$.
The terms denoted by capital letters correspond to the non-commutative operators with similar
symmetry properties of their tensor indices, 
\begin{equation}
{\hat{G}}_{ij} = {\hat{G}}_{ji} \, \ \ {\rm and} \, \ \  {\hat{F}}_{ij} = - {\hat{F}}_{ji}.
\label{twoparts}
\end{equation} 
Assuming that both non-standard effects, the asymmetry of the metric tensor and the non-commutative
character of its components are coming from the same quantum source whose manifestations become
detectable only on Planck's scale, it is natural to assume that the commutative part of $g_{ij}$
is symmetric, whether the skew-symmetric part is also a genuine non-commutative operator.
This leads us to the conclusion that it is enough to keep only two terms in the expression
\ref{gijgeneral}:
\begin{equation}
g_{ij} = g^0_{ij}  + {\hat{F}_{ij}},
\label{gijgeneral1}
\end{equation}
Here $g^0_{ij}$ also belongs to the non-commutative algebra of operators, but it remains in its
center and should commute with everything else. Having this point made clear, we shall drop the
distinctive upper signs and use the simplified notation $g_{ij} + F_{ij}$, remembering the
particular character of each term.

Thus the quantum theoretical arguments lead us naturally to the generalized form of metric
that served as the basic ingredient in the Born-Infeld generalization of electromagnetic lagrangian.

The generalized metric tensor should be now used in order to produce a new variational
principle defining the dynamical behavior of fields. We should follow the usual scheme
which consists in constructing the series of invariants of metric tensor starting 
from the lowest order which is the volume element. In classical differential geometry
one has:
\begin{equation}
\delta S = \delta \int \, \biggl[ \Lambda \, \sqrt{g} + \frac{1}{G} \, R \, \sqrt{g} + \gamma
\, ( R^{ijkm}R_{ijkm} - 4 R^ij R_ij + R^2) \, \sqrt{g} + ... \biggr]
\label{vargeneral}
\end{equation}
where $g$ denotes the determinant of $g_{ij}$ so that $\sqrt{g}$ defines the volume element in 
local basis; $R$ is the Riemann scalar, the next term is the Gauss-Bonnet invariant, and so forth.
The coefficients $\Lambda$, $1/G$ and $\gamma$ define the relative strength of each contribution.
We see now that the first term in the integrand has the form of the Born-Infeld lagrangian,
and is probably a good candidate for Planckian generalization of Einstein's cosmological term.
Taking into account the relative strengths of known classical fields, it is clear that the 
skew-symmetric tensor $F_{ij}$ appearing in the generalized metric does not describe 
the classical electromagnetic field, but rather the quantum field effects responsible 
for the early behavior of the Universe right after the Big-Bang. 

As in the Kaluza-Klein type theories, one can also expect the contribution of the $F$-field coming
from the next term containing the Riemann scalar; however, they enter only through their first and
second derivatives, as all other componenets of the metric tensor, therefore they will alter
only the derivative part of the lagrangian, and not the scalar potential part found in the
cosmological term.

In what follows, we shall expose in a concise way, on the example of the simplest finite
non-commutative algebra, which is the algebra of complex $n \times n$
matrices, how almost all the notions of usual differential geometry can be
extended to the non-commutative case. We shall also show how the gauge
theories and the analogs of the fibre bundle spaces and Kaluza-Klein 
geometries can be generalized in the non-commutative setting. The finite version of non-commutative
geometry is also the best adapted for the generalization of the determinant of metric 
tensor needed for the construction of the Born-Infeld lagrangian.

\subsection{The Non-Commutative Matrix Geometry}

\newcommand{\map}{\to}
\newcommand{\A}{\mathcal{A}}
\newcommand{\Mo}{\mathcal{M}}

We shall  generalize now the non-commutative Maxwell theory developed
in \cite{dubois-violette:90:II} in order to obtain  a Born-Infeld like theory.
Let us resume the notations and language of the theory. We consider  the algebra 
$\A = C^{\infty}(V) \otimes M_n(\
{C})$ with the
``vector fields'' spanned by the  derivations of $ C^{\infty}(V)$  and inner
derivations of $ M_n({\bf C})$. The differential algebra is
generated by the basis  of linear 1-forms acting on the derivations.
In order to construct a gauge theory, we must consider a finite projective module over $\A$.
By analogy with the Maxwell theory, we will consider the simplest one, \textit{i.e.} the free module of rank $1$ over $\A$, which can be identified with $\A$ itself.
Then one defines a gauge by the choice of a unitary element $e$ of $\A$, satisfying $h(e,e)=1$, with $h$ an hermitian structure on $\A$. 
Then any element of $\A$ can be written in the form $ e m$
with $m \in \A$ and  a connection on $\A$ is a map: 
\begin{equation}
  \nabla:  \A \map  \Omega^{1}(\A) ,  \ \, 
   e \ m \mapsto (\nabla e)\ m  + e \  dm
\end{equation}
In the gauge $e$, the connection can be completly characterized by an element
$\omega$ of $\Omega^{1}(\A)$:
\begin{align*}
\nabla e = e \  \omega \ .
\end{align*}
One can also  decompose $\omega$ in vertical  and horizontal parts:
\begin{equation}
\omega= \omega_h + \omega_v \, \ \ \, {\rm with} \, \ \ \, 
  \omega_h = A ,  \ \ \
  \omega_v = \theta + \phi
\end{equation}
Here $A$ is like the Yang-Mills connection, whereas $ \theta $ is the canonical 1-form of the matrix algebra, 
and plays the role
  of a  preferred origin in the affine space of vertical connections.
  It satisfies the equation:
  \begin{align*}
    d\theta +\theta^2=0
  \end{align*}
Then  $\phi$ is a tensorial form and can be  identified with scalar field
multiplet.

 Choosing a local basis of derivations of $\A$: $\{e_{\mu},e_{a}\}$,
 where for convenience $e_{\mu}$ are outer derivations of $C^{\infty}(V)$,
and $e_{a}= ad(\lambda_{a})$, with $\{\lambda_{a}\}$ a basis of anti-hermitian matrices 
of $M_n({\bf C})$, are inner derivations.\\
 The dual basis will be denoted by  $\{ \theta^{\mu}, \theta^{a} \}$.
 In this particular basis, we have:
 \begin{align*}
   A = A_{\mu} \theta^{\mu}\ , \
   \theta = - \lambda_a \theta^a\ , \
   \phi = \phi_a \theta^a 
 \end{align*}
If we choose the connection to be anti-hermitian, we can write   $ \phi = \phi_a^b \lambda_b \theta^a$.
The curvature tensor associated with $\omega$ is :
\begin{align*}
  \Omega = d\omega + \omega^2
\end{align*}
we can also define the field strength: 
\begin{align*}
  F&= dA + A^2 \ .
\end{align*}
Then by "dimensional reduction" one can identify:
\begin{align*}
\Omega_{\mu \nu} &= F_{\mu \nu} & \Omega_{\mu a} &= D_{\mu} \phi_a\\
\Omega_{a \mu } &= - D_{\mu} \phi_a &
\Omega_{a b} &= [\phi_a,\phi_b] - C_{a b }^{c} \phi_c
\end{align*}
where $ C_{a b }^{c} $ are the constant structure in the $\{\lambda_a\}$ basis.

A gauge transformation is performed by the choice of a unitary element $U$ of
$M_n({\bf C})$, satisfying $ h(e U, e U ) = 1$.
Then in the gauge $e'=e U$
\begin{align*}
  \omega' &= U^{-1} \omega U + U^{-1} dU
\end{align*}
$\theta$ is invariant under gauge transformations, then 
\begin{align*}
A'=U^{-1}  A U + U^{-1} dU  , \  \phi' =U^{-1}  \phi U 
\end{align*}

\subsection{The non-commutative version of the Born-Infeld lagrangian}

In this section, we will essentially recall the non-commutative generalization of the Born-Infeld lagrangian proposed in \cite{serie:04}.
By analogy with the abelian case, we want the lagrangian to satisfy the following properties: 
\vskip 0.2cm
\indent
1) One should find the usual NC Yang-Mills-Higgs theory in the limit $\beta \to \infty$
\vskip 0.2cm
\indent
2) The (non-abelian) analogue of the ``electric'' field strength should be bounded from above 
when the ``magnetic'' components vanish.
\vskip 0.2cm
\indent
(To satisfy this particular constraint, we must ensure that the polynomial expression under 
the root sign should start with terms
$1-\beta^{-2}(E^a)^2 + ...$ when $B^a =0$ ) 
\vskip 0.2cm
\indent
3) The action should be invariant under the action of the automorphisms of $\A$ (it includes diffeomorphisms and gauge transformations).
\vskip 0.2cm
\indent
4) The action must be  real.
\vskip 0.2cm
This enables us to introduce the following generalization of the Born-Infeld lagrangian density 
for a non-commutative gauge field:
\begin{align}
  \sqrt{\det|g|} - \{|\det ( {\bf 1} \otimes g + J \otimes \hat{\Omega} |\}^{1/4n}
\label{BI}
\end{align}
and $\hat{\Omega} = \Omega_{\alpha \beta}\hat{L^{\alpha \beta}} $ with
$\hat{L^{\alpha \beta}} $ the generators of the fundamental representation of
$SO(4+n^2-1)$,
$ \Omega_{\alpha \beta}$ are the components of the curvature defined in
previous section, and then are anti-hermitian elements of
$M(n,{\bf C})$.
In the expression above, $J$ denotes a unitary $2 \times 2$ matrix satisfying $J^2 = - {\bf 1}_2$, thus introducing a quasi-complex structure.This extra doubling of tensor space is necessary in order to ensure that the resulting lagrangian is real. 
At the same time, it raises the degree of the polynomial under the root up to $4 n$. 
We are left with the  root  of order  $4n$, so that the invariance of our action under the space-time diffeomorphism is preserve. 
\newline
\indent Let us recall a few arguments in favour of this construction:  
\newline
\indent 
The simplest way to generalize the Born-Infeld action principle in the framework of the non-commutative matrix geometry seems at first glance the substitution of real numbers by corresponding hermitian operators, like in quantum mechanics.  Then one would arrive at the following expression:
\begin{equation}
  \left\{\begin{array}{lcl}
       i F_{\mu \nu}  & \leftrightsquigarrow &  F_{\mu \nu}^a \otimes T_a \\      
      g_{\mu \nu}   & \leftrightsquigarrow & g_{\mu \nu} \otimes
      {\bf 1}_{n} \ ,\\      
    \end{array} \right.
  \label{correspondance}
\end{equation}
where $ {\bf 1}_{n}$ and $ iT_a$ are $n\times n$ hermitian matrices.  
What remains now to make the generalization complete, is to extend the notion of determinant 
taken over the space-time  indeces in the usual case, i.e. the determinant of a $4 \times 4$ matrix, to a notion of determinant taken in the tensor product of derivations and matrix indeces of the algebra $\A$. 
Then one would replace the objects in (\ref{BI1}) following the procedures in  (\ref{correspondance}) but it leads to a complex lagrangian.
In order to avoid this problem, it is necessary to start from an other form of the usual Born-Infeld lagrangian:
\begin{equation}\label{BI:J}
  S_{BI}[F,g] =\int_{{\bf R}^4}  \beta^2 \left( \sqrt{|g|} -
    \left| det_C \left( {\bf 1}_2 \otimes g_{\mu \nu} +\beta^{-1} \, \hat{J} \otimes
  i F_{\mu \nu} \,\right)\,\right|^{\frac{1}{4}} \right) d^4x \ ,
\end{equation}
where $\hat{J}$ is a $2 \times 2 $ unitary matrix whose square is equal to $- {\bf 1}_2$.
With the correspondence displayed in (\ref{correspondance}), we arrive at the action principle displayed in (\ref{BI}) which satisfies all the requirements we asked for,  1), 2), 3) and 4).
\\

The lagrangian in (\ref{BI}) contains the contribution of two types of fields:
the classical Yang-Mills potential, $ A = A_{\mu} \theta^{\mu}$, corresponding to the 
usual space-time components of the connection one-form, and the scalar multiplet coming 
from its matrix components $\phi = \phi_a \theta^a = \phi_a^b \lambda_b \theta^a$.  
In the case when $\phi =0$, this lagrangian coincide with the one studied in
\cite{serie:03}.
For cosmological considerations, we willrestrict ourselves to a qualitative analysis of the case
when the space time components of $\Omega$ do vanish $F_{\mu \nu}=0$, leaving
only the contribution of scalar multiplet degrees of freedom.

\subsection{The reduced lagrangian for scalar fields}

Let us recall the notations which will be used in the subsequent calculations.
The basis of matrix representation of the $SU(2)$-algebra is chosen as follows:

\begin{equation}
  \lambda_a = -i \sigma_a \, \ \ \, \ \
  \lambda_a \lambda_b = - \delta_{a b} + \sum_{c} \epsilon_{abc} \lambda_c \, \ \ \, \ \
  [\lambda_a, \lambda_b] = C_{ab}^{c} = 2 \epsilon_{abc} \lambda_c
\end{equation}

Now we have to evaluate the determinant of the following matrix:
\begin{align}
  \begin{vmatrix}
    1 & i D\hat{\phi}\\
    -iD\hat{\phi} & 1+i \hat{H} 
  \end{vmatrix}
  \label{matrice}
\end{align}
where 
 \begin{align}
   \hat{H} =  \left\{\hat{\Omega}_{ab} \right\}_{{a,b = 1,2,3}} \ , \
   D\hat{\phi}  = \left\{ D_{\mu}\hat{\phi}_a\right\}_{{a=1,2,3 \mu=0,1,2,3}}
 \end{align}
From now on, we choose the simplest ansatz\footnote{A detail analysis of other possible ansatzs is performed in \cite{serie-phd:05} } with one scalar field only:
\begin{align*}
 \phi = \varphi \ \theta
\end{align*}
After some algebra, we get the following result:

\begin{align}
  L=1- \left\{ 1+ 6 (D\varphi)^2 + 9
    (D\varphi)^4 + 16 \varphi^2(\varphi-1)^2 \right\}^{\frac{1}{4}} \sqrt{1+4 \varphi^2(\varphi-1)^2 }
\label{lagrangian-scalarfield} 
\end{align}
\indent
In \cite{serie:03} we have introduced a new non-abelian
generalization of the Born-Infeld  lagrangian, and found a family
of non-singular soliton-like solutions, using 't Hooft's ansatz
for the $SU(2)$ gauge potential.
In \cite{serie:04}, we have generalized this approach in order to
include scalar multiplets arising naturally in the non-commutative
geometry of matrix valued functions. In the  case when all degrees
of freedom are reduced to a single scalar field $\varphi$, we have
investigated homogeneous time dependent solutions. As expected,
the regular trajectories in the phase space $(\varphi,
\dot{\varphi})$ turn out to be comprised in a finite domain, thus
confirming the existence of a finite bound on field strength. This
particular behavior of the Born-Infeld like scalar field makes us
believe that it can be used in the framework of cosmological
models on a footing similar to the so called inflaton field.

This lagrangian will become the first ingredient for the cosmological model we want
to introduce. The next termin the lagrangian will be the usual Riemann scalar term providing the minimal interaction
with the scalar Born-Infeld field through covariant derivatives and through the
direct coupling with the symmetric part of metric tensor and interactions with the Yang-Mills fields.

\section{Non-commutrative Born-Infeld cosmology}

\subsection{General considerations}
\indent

The study of Born-Infeld lagrangians both in the abelian
\cite{breton} and the non-abelian \cite{galtsov1} case or as
condensates \cite{odintsov} has been widely developed in the last
few years in relation with string theory, where Born-Infeld type
lagrangians appear in a natural way
\cite{tseytlin:99,kerner-barbosa}.
Incidentally, scalar fields can be introduced in diffent ways in this context by considering the so-called Dirac-Born-Infeld (DBI) action wich could be relate to the usual Born-Infeld action by $T$-duality and dimensional reduction. 
Then the action we are considering can be seen as a non-commutative matrix DBI action.

 It has been argued that the $U(1)$ case shows the problem of the axial symmetry induced by the
electromagnetic field $F_{\mu\nu}$ which forbids  homogeneous
space-time solutions. On the other side in the framework of a non-abelian
gauge theory spherical symmetry can be restored and a
cosmological model in the standard FRW metric can be analyzed
\cite{kerner-barbosa,galtsov1}. In this case some interesting
results have been obtained. In fact it has been demonstrated that
this kind of lagrangian can furnish an inflationary scheme
\cite{galtsov1} or provide a theoretical framework to the phantom
field models \cite{odintsov} introduced in several theoretical attempts 
to explain the dark energy \cite{peebles:02,chaplygin02,shani:04,Bilic:2001cg}.\\
The introduction of a scalar multiplet, reduced to one
\cite{serie:04}, inside the $SU(2)$ gauge group, goes in this
direction, so that in the following we will consider a standard
FRW metric as the gravitational framework of our cosmological
model.
One may think that this kind of scalar field could drive the first stages of the evolution of the universe.


The most widely accepted scheme is to consider a scalar
field rolling down in its potential well slowly enough to
provide a cosmological constant like term in a first phase,
driving an exponential expansion. In a second phase the scalar
field decays in a true vacuum configuration oscillating coherently
around the minimum of its potential and transferring its energy to
the standard matter fields (reheating). Many approaches have been
proposed in time (old inflation, new inflation, \cite{liddle},
chaotic inflation \cite{linde}, see \cite{riotto} for a review)
also with completely different mechanism (Starobinsky scalaron
\cite{starobinski}).\\
The Born-Infeld scalar field may be interpreted just
as an effective bosonic field condensate and not a fundamental field.\\

In our model we shall consider minimal coupling with gravity,
obtained by adding the usual Einstein-Hibert lagrangian of
gravitational field, and by replacing in Euler-Lagrange equations
all derivatives by their covariant counterparts. The lagrangian
density will be

\begin{equation}\label{ncbi-lag0}
{\cal{L}} = {\cal{L}}_{ncbi}(\phi, \partial_{\mu}\phi, g_{\mu\nu})
+ \frac{1}{4\pi G}{R}\,.
\end{equation}

\noindent The Born-Infeld type scalar field lagrangian $
{\cal{L}}_{ncbi} $ contains two parameters: the maximal value of
the field strength $\beta$ and the characteristic mass $\gamma$ of
the scalar field. Together with Newton's constant $G$, our problem
will contain 3 explicit parameters with dimension of mass. By
letting $G \to 0$ the gravitational interaction is decoupled and
we recover the model of scalar Born-Infeld field studied recently
in \cite{serie:04}. We can also recover the usual $\phi^4$ theory
when the parameter $\beta$ tends to infinity. In what follows, a
systematic study of the dependence of the cosmological model on
the three aforementioned parameters will be presented.

\subsection{Cosmological model with Born-Infeld scalar field}\label{themodel}

The lagrangian (\ref{lagrangian-scalarfield}) can be write after a rescaling of the scalar field:

\begin{displaymath}
{\cal{L}}_{ncbi}=\beta^2\left\{1-\left[\left(1-{\beta}^{-2}
{\dot{\phi}}^2(t)\right)^2+\frac{16}{9}{\beta}^{-2}
\left(-\sqrt{3}\gamma+\phi(t)\right)^{2}\right]^{1/4}\right.
\end{displaymath}
\begin{equation}\label{ncbi-lag}
\left.\sqrt{1+\frac{4}{9}\beta^{-2}\left(-\sqrt{3}\gamma+
\phi(t)\right)^2\phi^2(t)}\right\}
\end{equation}

\noindent where the parameter $ \beta $ is the analog of the
BI cutoff for the non linear electromagnetic theory while $\gamma
$ gives account of the mass-shell of the scalar field and depends
from the non commutative algebra considered for the gauge group.
It is obvious that such a model contains in itself the $\phi^4$
theory \footnote{this can be observed while developing 
the expression (\ref{ncbi-lag}) in a  Taylor series as it will be shown in the next
section} so we expect to obtain a standard behavior when the
Born-Infeld parameter $\beta$ tends to infinity. To implement this
scheme in presence of gravity, the minimal coupling with the
Hilbert-Einstein lagrangian can be considered, with the action
given in the Eq.(\ref{ncbi-lag0}). It reads explicitly as follows:

\begin{equation}\label{ncbi-action}
{\cal{A}}=\int d^4x\,\,\,
\sqrt{-g}\frac{1}{{m_{Pl}}^2}R+\beta^2\left(1-\left(1+\Upsilon[\dot{\phi}(t),\phi(t)]\right)^{1/4}
\sqrt{1+\frac{4}{9}\Pi^2[\phi(t)]}\right)\,.
\end{equation}

The two functions $\Upsilon[\dot{\phi}(t),\phi(t)]$ and
$\Pi[\phi(t)]$ are defined as follows:

\begin{eqnarray}
\Upsilon[\dot{\phi}(t),\phi(t)]\,&=& \,
\left((1-\beta^{-2}\dot{\phi}^2(t))^2+\frac{16}{9}\Pi^2[(\phi(t)]\right)\\
\Pi[\phi(t)]&=& \beta ^{-1}\phi(t)(\phi(t)-\sqrt{3}\gamma)\,.
\end{eqnarray}

Varying (\ref{ncbi-action}) with respect to the 
metric  we get two cosmological equations:

\begin{displaymath}
\frac{\dot{a}^2(t)}{a^2(t)}\,+\,\frac{k}{a^2(t)}
=\frac{1}{3}\frac{\beta^2}{m_{Pl}^2}\left\{\beta^{-2}{\dot{\phi}^2(t)}(1-\beta^{-2}{\dot{\phi}^2(t)})
\Upsilon^{-\frac{3}{4}}[\dot{\phi}(t),\phi(t)]\left(1+\frac{4}{9}\,\Pi^2[\phi(t)]\right)^{1/2}
+\right.
\end{displaymath}
\begin{equation}\label{ncbi-eq1}
\left.{\Upsilon^{\frac{1}{4}}[\dot{\phi}(t),\phi(t)]}\left(1+\frac{4}{9}\,\Pi^2[\phi(t)]\right)^{1/2}-1\right\}\,,
\end{equation}

which is the energy equation (Friedmannn equation) and
\begin{equation}\label{ncbi-eq2}
2\frac{\ddot{a}(t)}{a(t)}\,+\,\frac{\dot{a}^2(t)}{a^2(t)}\,+\,\frac{k}{a^2(t)}
=-\frac{\beta^2}{m_{Pl}^2}\left\{1-\Upsilon^{\frac{1}{4}}[\dot{\phi}(t),\phi(t)]\right\},
\end{equation}
which is the pressure equation. 

The third equation is obtained varying the lagrangian with respect to the scalar
field. It can be shown that it coincides with the
conservation equation for the energy momentum tensor, namely the
Bianchi identity \cite{deritis}. We have:

\begin{displaymath}
\ddot{\phi}(t)\left(1+\frac{4}{9}\Pi^2[\phi(t)]\right)\left\{\left(1-\beta^{-2}\dot{\phi}^2(t)\right)^2+
\frac{16}{9}\Pi^2[\phi(t)]\left(1-3\beta^{-2}\dot{\phi}^2(t)\right)\right\}+
\end{displaymath}
\begin{equation}\label{kl-gordon-gen}
+\frac{4}{9}\phi(t)(\phi(t)-\sqrt{3}\gamma)
(2\phi(t)-\sqrt{3}\gamma)\left(\Upsilon[\dot{\phi}(t),\phi(t)]\right)\times
\end{equation}
\begin{displaymath}
\times\left(3-\beta^{-2}\dot{\phi}^2(t)+
\frac{8}{3}\Pi^2[\phi(t)]\right)
-6\beta^{-2}\dot{\phi}^{2}(t)\left(1-\beta^{-2}\dot{\phi}^2(t)\right)\left(1+
\frac{4}{9}\Pi^2[\phi(t)]\right)+
\end{displaymath}
\begin{displaymath}
+3\frac{\dot{a}(t)}{a(t)}\,\dot{\phi}(t)\left(1+\frac{4}{9}\Pi^2[\phi(t)]\right)
\left(1-\beta^{-2}\dot{\phi}^2(t)\right)\left(1-2\beta^{-2}\dot{\phi}^2(t)+\beta^{-4}\dot{\phi}^4(t)
+\frac{16}{9}\Pi^2[\phi(t)]\right)\,.
\end{displaymath}

\noindent From the lagrangian or directly from the cosmological
Eq.(\ref{ncbi-eq1}) and Eq.(\ref{ncbi-eq2}) we find that the
Higgs-type scalar field is characterized by its energy density and
the pressure term:

\begin{equation}
\rho_{ncbi}=\beta^2\left\{\beta^{-2}{\dot{\phi}^2(t)}(1-\beta^{-2}{\dot{\phi}^2(t)})
\Upsilon^{-\frac{3}{4}}[\dot{\phi}(t),\phi(t)]\left(1+\frac{4}{9}\,\Pi^2[\phi(t)]\right)^2
+{\Upsilon^{\frac{1}{4}}[\dot{\phi}(t),\phi(t)]}-1\right\}\,,
\label{energy-dens}
\end{equation}

\begin{equation}
p_{ncbi}=
\beta^2\left\{1-\Upsilon^{\frac{1}{4}}[\dot{\phi}(t),\phi(t)]\right\}\,,
\label{pressure}
\end{equation}

which allows us to write the barotropic factor as well

\begin{equation}
w_{ncbi}=\frac{\rho_{ncbi}}{p_{ncbi}}\,.
\end{equation}

At the first sight the system appears
quite complex. This is why, as a first step in what follows, we will
perform a qualitative analysis of the model through numerical
study of the solutions. In this way it will be possible to get
significant conclusions about the dynamics of the system and
its cosmological interpretation.

\subsection{The Born-Infeld relaxed limit}

We have seen that in the general case the NCBI lagrangian
(\ref{ncbi-lag}) is characterized by means of two parameters like
in the standard Higgs theory. However in this case their origin 
is strictly connected with the non-abelian character of the gauge field
and with the Born-Infeld model itself.

One of these parameters, namely $\gamma$, comes from the non
abelian character of the Yang-Mills theory and it is connected
with the scaling of the non commutative algebra.
\cite{serie:03, serie:04}. On the other side, the parameter $\beta$
arises in relation to the choice of a Born-Infeld type lagrangian
for the gauge theory \cite{kerner-barbosa}. It acts as the
Born-Infeld typical threshold for the field intensity. The two
parameters have different dimensions. In our formalism the NC
Yang-Mills gauge field $\Omega_{\mu\nu}$  which also contains the scalar field 
\cite{serie:03,serie:04} has the dimensions of
$[cm]^{-1}$ as the scalar field itself and $\gamma$.
This result indicates that one can consider $\gamma$ as
a mass-shell parameter of the scalar field imitating the typical
mass parameter of the Higgs field. Regarding  $\beta$ it is easy
to check that 
(\ref{ncbi-lag}) it has to be $[cm]^{-2}$.

Now, to compare with the linear limit of the NCBI
lagrangian, we can develop it in series of $\beta$. As expected,
the NCBI model shows, in the linear limit, a typical Higgs
behavior with a ``top hat" potential:

\begin{equation}\label{top-hat}
{\cal L}_{ncbi}^{\beta\rightarrow{\infty}}=
\frac{1}{2}\dot{\phi}^2(t)-\sqrt{2}\phi^2(t)\left(\frac{\sqrt{3}}{3}\phi^2(t)-\gamma\right)^2
+o[\frac{1}{\beta}]^2\,\,.
\end{equation}

\noindent The shape of the potential depends only on the
coefficient $\gamma $, which sets of the mass of
the scalar field. Of course the analogy with the Higgs field is
not complete because of the Born-Infeld linear limit has no
coupling constant to compare with the usual Higgs sector. 
The different choices of the values of parameters $\beta$ and $\gamma$
will generate not only a whole spetrum of inflationary behaviors of our model,
but will also create a possibility for the acceleration as well.

\section{The phase space study}

\subsection{Preliminary Considerations}

The field equations of the NCBI model appear extremely complicated and there is little hope 
to solve them analytically. But qualitative analysis can be performed instead. To do this we should analyze 
the phase space of the system considering simultaneous evolution of the scale factor and of the scalar field.
We assume a flat space-time ($k=0$ in Eq. (\ref{ncbi-eq1})-(\ref{ncbi-eq2})),
since more and more results coming from the CMBR observations seem to confirm the spatially flat model. 
It also seems that this assumption is well supported by other astrophysical data. After the COBE results of 1992 
\cite{cobe} other new data have been obtained in the last years. The balloon-based experiments \cite{cmbr-palloni} 
and the 2003 the WMAP satellite provided stringent estimates of the last scattering surface anisotropies
indicating that the actual spatial curvature is very close to zero.

A few more considerations will enable us to better understand the physical properties of the NCBI model. Let us
start with the analysis of the peculiarities of the NCBI fluid. Looking at the definitions of $\rho_{ncbi}$ and
$p_{ncbi}$ (\ref{energy-dens}), (\ref{pressure}) we can observe that the relation between these quantities can
be recast in a more significant form:

\begin{equation}\label{press-enedens}
p_{ncbi}\,=\,\rho_{ncbi}^c-\rho_{ncbi}\,,
\end{equation}

\noindent introducing a sort of critical $\rho_{ncbi}^{c}$ energy density of the model defined as:

\begin{equation}\label{energy-dens-crit}
\rho_{ncbi}^c\,=\,\beta^2\left\{\beta^{-2}{\dot{\phi}^2(t)}(1-\beta^{-2}{\dot{\phi}^2(t)})
\Upsilon^{-\frac{3}{4}}[\dot{\phi},\phi]\left(1+\frac{4}{9}\,\Pi^2[\phi]\right)^\frac{1}{2}\right\}\,.
\end{equation}

Now, combining with the energy density, the relation (\ref{press-enedens}) can be written as follows:

\begin{equation}\label{press-enedens2}
p_{ncbi}\,=\,\left(w_{ncbi}^{crit}-1\right)\rho_{ncbi}\,,
\end{equation}

\noindent with $w_{ncbi}^{crit}\,=\,\frac{\rho_{ncbi}^c}{\rho_{ncbi}}$. This definition
defines the properties of the NCBI fluid.

One checks easily that the energy density can take on negative values which will lead to the 
so-called ghost solutions. Taking into account the regions of the phase space $\{\phi,\dot\phi\}$ where the
energy density has only positive values we can deduce the conditions
on the equation of state of the fluid in relation to the value of $w_{ncbi}^{crit}$. In fact, depending on the
sign of this quantity, we shall get different situations as shown in the following table \ref{table-w}:

\begin{table}[hbt]
\begin{center}
\begin{tabular}{|c|c|c|c|}
  \hline
  $w_{ncbi}^{crit}$ & $w_{ncbi}$ & Cosmological behavior & $\dot{\phi}$ \\
  \hline
  \hline
 $>\,0$ & $]-1\,,\ 0\,[$ & Inflaton-like  & $]-\,\beta\,,\ \beta\,[$\\
  \hline
   $0$ & $-1$ & Cosmological constant-like  & $\beta$\\
   \hline
    $<\,0$ & $<-1$ & Phantom-like  & $]-\infty \,,\ -\beta\,[\, \bigcup \, ]\,\beta\,,\ +\infty [$\\
    \hline
\end{tabular}\caption{\label{table-w} {\small{The cosmological behavior depending on the equation 
of state of the fluid.}}}
\end{center}
\end{table}
\noindent These results can be easily recovered looking to the definition (\ref{press-enedens2}). In the last
column are given the constraints on the scalar field defining the different behaviors. From
the definition of $\rho_{ncbi}^c$ it is easy to observe that the sign of this term and then as a
consequence, the positivity energy density, of $w_{ncbi}^{crit}$ is ruled by the term
$(1-\beta^{-2}{\dot{\phi}^2(t)})$. As a consequence, the Inflaton-like rate, the Phantom effect and
the cosmological constant-like behavior are obtained if respectively  $\dot\phi$ is between $-\beta$ and
$\beta$, when this quantity is out of this interval or if $\dot\phi$ is equal to the NCBI parameter itself.

Let us now discuss the behavior of this model with respect to the various Energy Conditions. Again, it
will be possible to provide only qualitative and numerical calculations, but they will allow us to deduce some
interesting properties of the model.
We know that Energy Conditions can take on several forms. Each of them will define different regions
in the phase space of the scalar field.

\begin{figure}[htbp]
\begin{center}
  \includegraphics[width=12cm, height=7.5cm]{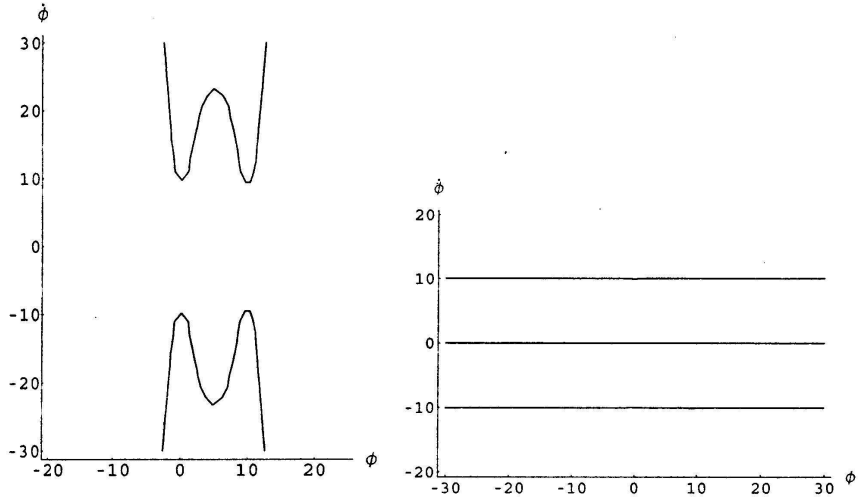}
 \caption{\label{fig-encond1} {\small{This graph shows the allowed region in the plane $\{\phi(t),\ \dot{\phi}(t)\}$ 
defined by the Weak Energy
 Condition. The plot is obtained in the case of $m_{pl}\,=\,10$, $\beta\,=\,10^{-1}m_{pl}$ and $\gamma\,\sim\,m_{pl}$.
 The left part shows the energy density curve, the regions inside the curve correpond to negative values, which
become positive outside. The right panel shows the condition $\rho_{ncbi}+p_{ncbi}\,>\,0$, which are positive
 between the two straight lines and negative outside. The $\dot\phi\,=\,0$ line gives a
 cosmological constant behavior as implied by the Eq. (\ref{energy-dens-crit}).}}}
\end{center}
\end{figure}

The positivity of the energy density parameter is important since it prevents the system from possible vacuum
instability and rules out the ghost-type solutions. Imposing such a condition selects a
particular region in the phase space, see Fig. \ref{fig-encond1}. Moreover, the Weak Energy Condition
$\rho_{ncbi}\,\geq\,0$ and $\rho_{ncbi}\,+\,p_{ncbi}\geq\,0$, is strictly related to the dynamics of the field
since the second relation directly affects the value of $w_{ncbi}$ and rules out supraluminal solutions.

\begin{figure}[htbp]
\begin{center}
  \includegraphics[width=11cm, height=6cm]{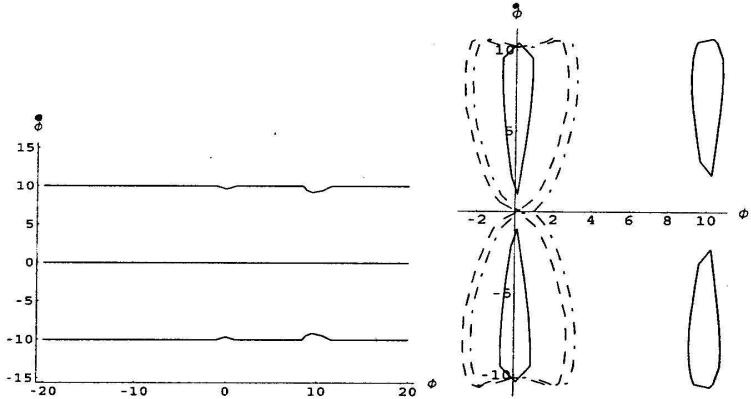}
 \caption{\label{fig-encond2}  {\small{The plot showing the domain in the phase space satisfying the 
Dominant Energy Condition (left) and the Strong Energy Condition (right). The plot corresponds to the choice of
$m_{pl}\,=\,10$, $\beta\,=\,10^{-1}m_{pl}$. The left graphic is not influenced by the choice of different
values for the parameter $\gamma$ if $\beta$ is set.h e Strong Energy Condition modifies the
contour plots. The three different curves in the right part refer to: $\gamma\,\sim\,m_{pl}$ (solid),
$\gamma\,\sim\,10^{-1}m_{pl}$ (dashed) and and $\gamma\,\sim\,10^{-2}m_{pl}$ (dot-dashed). The inner regions 
correspond to positive energy values while the external ones are negative and provide suitable initial condition 
for accelerating dynamics.}}}
\end{center}
\end{figure}

The Dominant Energy Condition and the Strong Energy Condition determine some areas of the scalar field
phase space $\{ \phi, \, \dot{\phi} \}$ within which these conditions are satisfied, see Fig. \ref{fig-encond2}. 
In particular the values
of $\phi\,,\ \dot\phi$ violating the Strong Energy Conditions lead to accelerating solutions.

The results are plotted in Fig. \ref{fig-encond1} and Fig. \ref{fig-encond2} for a particular choice of
values of parameters $\beta$ and $\gamma$, but these plots can be easily generalized to other values of this
quantities yielding similar or slightly different shapes. The results obtained from the
imposition of the Strong Energy Condition are interesting, too. In such a case the positive value regions are 
defined around certain values of the scalar field, with large areas outside where this condition is violated. This
implies that in order to achieve accelerated regime in the later stages
the scalar field does not necessarily need any fine tuning of initial conditions.

\subsection{Critical points}

Let us now investigate the dynamical properties of the NCBI scalar field. In terms of relevant variables only, the system evolves in a four dimensional space ($a(t),\dot{a}(t), \phi, \dot{\phi}(t)$) which can be reduced to a three-dimensional set. In fact, in a spatially flat
case it is possible to introduce the Hubble parameter as an independent variable and to study the three-dimensional
phase space ($H, \phi, \dot{\phi}$) (here and in the following we neglect the explicit time dependence of 
dynamical variables). Furthermore, the Friedmann equation provides a surface in this three
dimensional space,  implying that the relevant dynamics can be followed on this two-dimensional surface.

However, due to the complexity of the model, it is impossible to get an explicit
expression of both $H$ and $\dot{\phi}$. To overcome this problem, a reasonable approach is to solve 
the field equations numerically and to reconstruct the phase space curves for several values of parameters.

The first thing to do while performing a qualitative study is to find the critical points of the system. The
critical points of the scalar field can be obtained by solving Eq.(\ref{kl-gordon-gen}) with $\dot{\phi}=0\,\,
\text{and}\,\, \ddot{\phi}(t)=0$, they are obtained the three values $\phi= 0,\ \gamma\sqrt{3}/2\  \text{or}\
\gamma\sqrt{3}$. On the other side, the critical points of the full system are obtained considering also the
other field equations. From the cosmological point of view they correspond to de Sitter solutions with 
constant scalar field. The general situation can be summarized as follows:

\begin{eqnarray}\label{critical-points}
\left(H,\phi\right)\,=\,\left\{\begin{array}{lr}
\ (0,0)\ \ \ \ \ \ \ \ \ \ \mbox{Minkowski}\,,\\
\\\left(\left(\frac{1}{3}\frac{\beta^2}{m^2_{Pl}}\left(\left(1+\beta^{-2}\gamma^4\right)^\frac{1}{2}
\left(1+\frac{\beta^{-2}\gamma^4}{4}\right)^\frac{1}{4}\right)-1\right)^\frac{1}{2},\frac{\gamma\sqrt{3}}{2}\right)\
\ \mbox{de Sitter}\,,\\ \\ \ \left(0,\gamma\sqrt{3}\right)\ \ \ \ \ \ \ \,\mbox{Minkowski}\,.\end{array}\right.
\end{eqnarray}

\noindent The first and the third fixed points correspond to trivial Minkowski spaces \cite{gunzig}. In fact
from the definition of the energy density (\ref{energy-dens}) and pressure (\ref{pressure}) we see
that in these cases both do vanish. In other words there is no source for dynamics and the
system reduces to a Minkowski space-time.

The second fixed point is a typical de Sitter solution with the equation of state for a scalar field of the
cosmological constant type. The energy density is different from zero and provides a source for the
exponential expansion:

\begin{equation}
\rho_{ncbi}^{deSitter}={\beta^2}\left\{\left(1+\beta^{-2}\gamma^4\right)^{1/2}
\left(1+\frac{\beta^{-2}\gamma^4}{4}\right)^{1/4}-1\right\}.
\end{equation}

As we will see later, in our model this solution is no more an attractor in the usual sense. In fact, there
are some values of the parameters for which phase trajectories deviate from the de Sitter solution even if
the initial conditions are taken very close to this fixed point. This result does not agree with the
 the No-Hair Theorem \cite{kolb} which holds for general cosmological models containing 
scalar field, as demonstrated in \cite{gunzig} both with a minimal and a non-minimal coupling cases.
Similar results concerning the dynamics of a Born-Infeld scalar field system have been recently obtained
in \cite{novello05}. 


The interpretation of the scalar field fixed points becomes obvious with the Born-Infeld relaxed limit
analyzed in the previous section. It is easy to check that the two values $\phi(t)= (0, \sqrt{3}\gamma)$
correspond to the true vacuum configuration of the top hat potential in Eq.(\ref{top-hat}) while $\phi(t)=
\sqrt{3}/2 \gamma$ corresponds to the unstable ``false vacuum" state. In what follows, we
shall sometimes refer to these states as with the usual double-well Higgs potential.
Let us now analyze the stability of the fixed points. We can rewrite the NCBI lagrangian (\ref{ncbi-lag}) in a
new form and rescale the scalar field by substituting $\phi \rightarrow \sqrt{3} \phi$:

\begin{align*}
{\cal{L}}_{ncbi}&=\beta\left[
 1-{\left( {\left( 1 -\frac{3\,{\dot{\phi} }^2}{{\beta }^2} \right) }^2 + \frac{16\,{\phi }^2\, 
{\left( -\gamma  + \phi  \right) }^2}{{\beta}^2} \right) }^{\frac{1}{4}}\left(1 + \frac{4\,{\phi }^2\,{\left( -\gamma  + \phi  \right)
 }^2}{{\beta}^2}\right)^{\frac{1}{2}}\right]
\end{align*}

In the vicinity of singular points  we observe behaviors similar to those of a scalar field minimally
coupled to gravity \cite{belinsky:85}, with energy density given by:

\begin{equation}
  \rho=b(\frac{1}{2}\dot{\phi}^{2} +\frac{1}{2}m^{2} \phi^{2} ) + \rho_{0}\,.
\end{equation}
\noindent
For each fixed point, the parameters $m^{2}$ and $\rho_{0}$ should be expressed in terms of the parameters
$\beta$ and $\gamma$ (the parameter $b$ is not relevant for the stability analysis). We just saw that for
the ``Born-Infeld scalar field''  there are the three fixed points (\ref{critical-points}), which in terms of
the scalar field variables $\phi, \dot\phi$ are given by $(\phi=0,\ \dot\phi=0)$, $(\phi=\gamma,\ \dot\phi=0)$
and $(\phi=\frac{\gamma}{2},\ \dot\phi=0)$.

Now, linearizing the equations of motion around these fixed points and
evaluating the corresponding Jacobian matrices, we infer the exact character of each singular point \cite{libro-sist-din}. 
For the two asymptotic Minkowskian solutions, that
is at the points $(\phi=0,\ \dot\phi=0)$ and $(\phi=\gamma,\ \dot\phi=0)$, we find that

\begin{equation}
 m^{2}=4\gamma^{2} \ \ \ \ \  \ \ \ \  \rho_{0}=0\,.
\end{equation}

\noindent which corresponds to a stable point with eigenvalues in the $(\phi,\ \dot\phi)$ plane: $\lambda_{{\pm}}=
\pm i 2\gamma$, both complex. In the case of the point $(\phi=\frac{\gamma}{2},\ \dot\phi=0)$ we find

\begin{equation}
m^{2}\,=\,-2\gamma^{2} \frac{(1+\displaystyle\frac{\gamma^{4}}{2\beta^{2}})}{(1+\displaystyle\frac{\gamma^{4}}{4\beta^{2}})} 
\, \ \ \, \ \  
 \rho_{0}\, =\,\left(1+\frac{\gamma^{4}}{\beta^{2}}\right)^{\frac{1}{4}} \sqrt{1+\frac{\gamma^{4}}{4\beta^{2}}}  - \,1\,.
\end{equation}

Hence, because $m^{2}$ is negative, it corresponds to an unstable point, and the eigenvalues in the plane
$(\phi,u)$ are:

\begin{equation}
  \lambda_{+}\,=\,\displaystyle\frac{3H_{0}}{2} ( \sqrt{1-\frac{4m^{2}}{9H_{0}^{2}}}-1)  > 0 \, \ \ \, \ \
  \lambda_{-}\,=\,-\displaystyle\frac{3H_{0}}{2}(1 - \sqrt{1-\frac{4m^{2}}{9H_{0}^{2}}})  < 0
\end{equation}

\noindent with $H_{0}= \sqrt{\frac{\kappa^2}{3}\rho_{0}}$. In terms of $\beta$ and $\gamma$, the eigenvalues
take on the following form:
  \begin{multline}
    \lambda_{+}\,=\, \frac{3}{2\sqrt{6}}\, {\sqrt{-2 + {\left( 1 +
            \frac{{\gamma }^4}{{\beta }^2} \right) }^{\frac{1}{4}}\,
        {\sqrt{4 + \frac{{\gamma }^4}{{\beta }^2}}}}} \ \times\\
    \left( -1 + \,{\sqrt{1 + \frac{32\,{\gamma }^2\,\left( 2\,{\beta
              }^2 + {\gamma }^4 \right) } {3\,\left( 4\,{\beta }^2 +
              {\gamma }^4 \right) \, \left( -2 + {\left( 1 +
                  \frac{{\gamma }^4}{{\beta }^2} \right)
              }^{\frac{1}{4}}\, {\sqrt{4 + \frac{{\gamma }^4}{{\beta
                    }^2}}} \right) }}} \right)
  \end{multline}
\begin{multline}
\lambda_{-}\,=\, \frac{-3}{2\sqrt{6}}
 {\sqrt{-2 + {\left( 1 + \frac{{\gamma }^4}{{\beta }^2} \right) }^{\frac{1}{4}}\,
           {\sqrt{4 + \frac{{\gamma }^4}{{\beta }^2}}}}}\, \times \\
      \left( 1 + {\sqrt{1 + \frac{32\,{\gamma }^2\,\left( 2\,{\beta }^2 + {\gamma }^4 \right) }
              {3\,\left( 4\,{\beta }^2 + {\gamma }^4 \right) \,
                \left( -2 + {\left( 1 + \frac{{\gamma }^4}{{\beta }^2} \right) }^{\frac{1}{4}}\,
                   {\sqrt{4 + \frac{{\gamma }^4}{{\beta }^2}}} \right) }}} \right) 
\end{multline}

The most interesting feature for the stability study is the sign of $\lambda_{+}$ and $\lambda_{-}$
which is not easy to determine. However, after the expansion in powers of $\beta$ for
$\lambda_{\pm}$ these expressions are considerably simplified:

\begin{equation}
  \lambda_{+}\,=\,\displaystyle\sqrt{2}\gamma - \frac{3}{4}\sqrt{\frac{\gamma^{4}}{2\beta^{2}}} + o(\frac{1}{\beta^{2}}) \, \ \ \, \\ \,
  \lambda_{-}\,=\,-\displaystyle\sqrt{2}\gamma - \frac{3}{4}\sqrt{\frac{\gamma^{4}}{2\beta^{2}}} + o(\frac{1}{\beta^{2}})\,.
\end{equation}

\subsection{Numerical Analysis}

Our dynamical system is characterized by three parameters $\beta,\ \gamma,\ m_{Pl}$, which in
the case of flat space geometry determine it completely. However it is obvious that only two of them
are independent. To perform the numerical analysis we have considered $\beta, \gamma$ in terms of the normalized
Planck mass. In order to explore a dimensionally uniform parameter space we consider $1/\beta, 1/\gamma^2, 1/m^2_{Pl}$
\footnote{We remember that the Planck mass has the
dimension of $[cm]^{-1}$, while the Born-Infeld parameter has the dimension of $[cm]^{-2}$ and the mass
parameter $\gamma$ has the dimension of $[cm]^{-1}$.}.

The parameters $\beta$ and $ \gamma$ cannot take on arbitrary values if we have to satisfy the following physical
requirements: 
\begin{itemize} \item[$\, i)$] the characteristic Born-Infeld parameter cannot be higher than the squared Planck
length, i.e. $\beta \lesssim{m^2_{Pl}}$; \item[$\, ii)$] the mass parameter also can not be greater than the Planck
mass, i.e. $\gamma^2\lesssim{m^2_{Pl}}$. \end{itemize} Taking into account these constraints we have explored
the space of the parameters considering a domain ranging from fractions of the Planck constant to its entire
value. To perform a numerical analysis we have fixed the value of $m_{Pl}$ at 10, so that the essential features 
of the phase space behavior can be clearly displayed. To give a picture of the results we have considered 
$\beta\in [10^{-2}m^2_{Pl}\,,m^2_{Pl}]$ (namely with $m_{Pl}=10$, $1/\beta \in [10^{-2}, 1]$) and 
$\gamma\in [10^{-2}m_{Pl}\,,m_{Pl}]$ ($1/\gamma^2 \in [10^{-2}, 100]$).

The definition of the lagrangian and the field equations impose some
natural constraints in the phase space  $\{\phi\ ,\dot{\phi}\}$. The lagrangian (\ref{ncbi-lag}) 
contains an even root term, this means that the term under the root must be positive. Indeed it is
as it is a sum  of two squared terms. Let us look at the field
equations. The Friedmannn equation (\ref{ncbi-eq1}), in the spatially flat case we are considering, implies the
positivity of energy density $\rho_{ncbi}\geq{0}$. This gives the
first constraint on the phase space of the scalar field as we already saw in Fig.\ref{fig-encond1}.
Let us also note that there are no divergences in the definition of the energy
density and that it remains always finite.

Considering now the pressure equation (\ref{ncbi-eq2}) we observe that the only constraint is again the
positivity of the function under root, but it leads to the same constraints as those imposed by the
lagrangian itself.

The analysis of the scalar field equation (\ref{kl-gordon-gen}) is more complex. Here we 
observe that if the function multiplying the $\ddot{\phi}(t)$ term is zero than the term
$\ddot{\phi}(t)$ can become divergent. This implies, as in the case without gravity \cite{serie:04}, that there exists
a characteristic curve in the phase space where the solutions for the scalar field become divergent in a
finite time. This problem can be cured if all other terms in the equation simultaneously vanish. This
allows $\ddot{\phi}(t)$ to remain finite and provides some crossing points in the phase space on the
separatrix were the system can go beyond this curve. The singular curve 'separatrix) is given by the 
following relation:

\begin{equation}
\left(1+\frac{4}{9}\Pi^2[\phi(t)]\right)\left\{\left(1-\beta^{-2}\dot{\phi}^2(t)\right)^2+
\frac{16}{9}\Pi^2[\phi(t)]\left(1-3\beta^{-2}\dot{\phi}^2(t)\right)\right\} = 0\,.
\end{equation}

\noindent Now, comparing with the pure Higgs field case studied in \cite{serie:04} there is an obvious and
important difference. The presence of gravity, which induces the new constraint $\rho_{ncbi}\geq 0$
implies the absence of crossing points over the divergence curve.


With the allowed regions thus defined, the following step is to plot the characteristic curves generated by
the scalar field equation. The cosmological evolution of the model and the scale factor behavior are driven
by the scalar field dynamics.

At this point we are ready to perform a numerical study of the system.  In the following
we will present the results for three combinations of parameters expressed in terms of the Planck mass:\\ \\
Case: $\beta=m^2_{Pl}$.\\
The system displays similar phase trajectories for all values of mass parameter (we plot the case
$\gamma \sim m_{Pl}$ and $\gamma\sim 10^{-1}m_{Pl}$). The scalar field shows a spiralling behavior
ending up in one of the two stable fixed points.

\begin{figure}[htbp]
\begin{center}
\includegraphics[width=7cm,height=6cm]{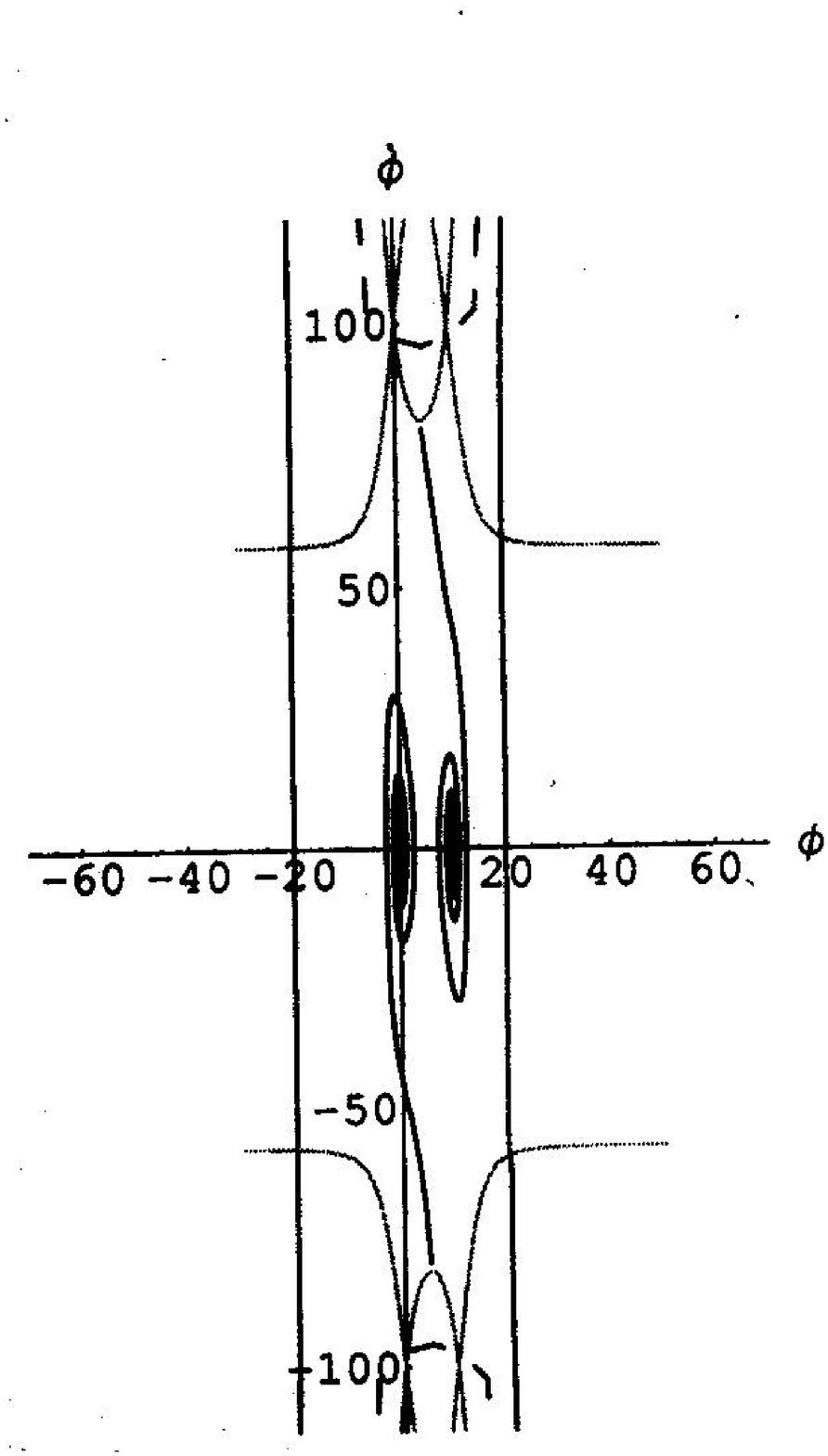}
\includegraphics[width=4cm,height=5.5cm]{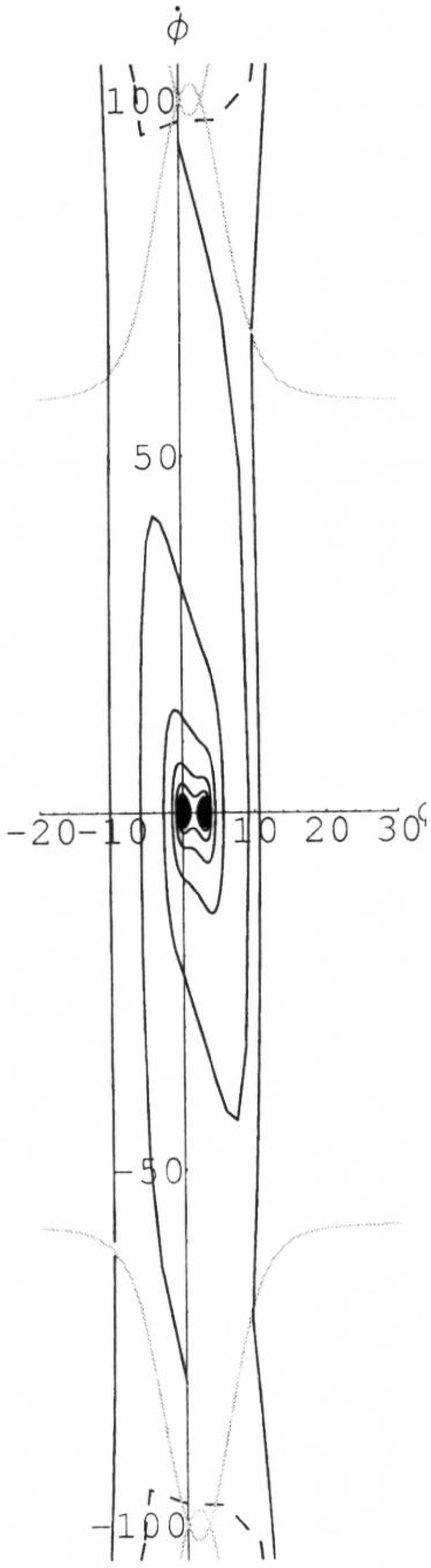}
\caption{\label{plot1} {\small{Phase space of the scalar field in the case $\beta=m^2_{Pl}$ with $\gamma \sim m_{Pl}$
and $\gamma \sim 10^{-1}m_{Pl}$. The dashed curve is the constraint obtained by the Weak Energy Condition (as
yet seen in Fig.\ref{fig-encond1}), while the lighter one is the characteristic curve due to the scalar field
equation. We show only some characteristic curves since for other initial conditions the shape is analogous.}}}
\end{center}
\end{figure}

The time evolution on phase trajectories is from left to right. For smaller values of $\gamma$ the allowed
region around the fixed points becomes narrower. This is understandable because the mass parameter in the standard
$\phi^4(t)$ theory controls the width of the potential. One obvious difference with the case analyzed in
\cite{serie:04} is the spiraling behavior. This effect is due to the presence of a friction term in the
scalar field equation coming from the gravitational interaction. The pure Born-Infeld regime is recovered when
one increases the values of the Planck mass (equivalent with a decrease of gravitational coupling) and fixes the
other parameters. On the other side if one increases the $\beta$ parameter the effect is to widen the 
allowed domain along the $\dot{\phi}(t)$ direction. This transforms the system into the standard Higgs case in
presence of gravity obtained in the relaxed limit (\ref{ncbi-lag0}).

A particular result visible in Fig.(\ref{plot1}) is that the physically significant dynamics is constrained
to a very small region around the fixed points. All the trajectories with different initial conditions
intersect the singular curve in a finite time. This entails a singular behavior for the scale factor thus
excluding such solutions. It is easy to find the cases in which it is possible to get
inflationary behavior. In general (see Tab.\ref{table}) non-linearities combined with a small enough
 mass parameter make the
scalar field roll down too quickly, never attaining a slow rolling regime. This implies that the scale factor follows
a power law regime since the beginning.

This conclusion results from examining the slow rolling parameter $\varepsilon =
-\frac{\dot{H}(t)}{H^2(t)}$ \cite{riotto}. It is well known that this parameter should be
significatively smaller than 1 throughout all the inflationary expansion.\\ 
In the cases $\gamma\sim\,m_{Pl}$,
$\gamma\sim\,10^{-1}m_{Pl}$ and $\gamma \sim\,10^{-1}m_{Pl}$ with a slow rolling-type initial condition from the
top of the unstable fixed point, it is possible to get inflationary evolution with
the right power law phase transition at a given stage.\\ 
Case: $\beta=m^2_{Pl}/10$.\\
We examine the plot with $\gamma$ of the same order of the Planck mass. This plot is quite interesting since it
is clear that the two stable fixed points define two separate regions between which there are no trajectories.
The other cases with smaller values of the mass parameter recall quite similar shapes to the previous case.

\begin{figure}[htbp]
\begin{center}
\includegraphics[width=8.1cm,height=4.7cm]{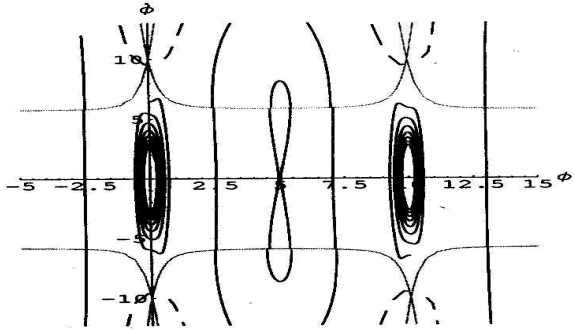}
\caption{\label{plot2-1} {\small{Phase space of the scalar field in the case $\beta=,m^2_{Pl}/10$ with
$\gamma\sim\,m_{Pl}$. The dashed curve is the constraint imposed by the Friedmann equation, while
the thinner one is due to the scalar field equation. We limit the number of solutions 
plotted to make the results clearly visible.}}}
\end{center}
\end{figure}

\noindent Around the two fixed points there are stable spiralling solutions. The trajectories starting from the
unstable fixed point hit the singular curve in a finite time. They describe closed lines broken
by the characteristic curve. If the initial velocity is small enough, the scalar field remains in the unstable critical
point for a long time and then it joins again the characteristic curve in finite time since the remaining part of
the phase space is forbidden. This provides an infinite inflationary solution of cosmological constant
type. It is obvious that in this case the de Sitter solution is no more an attractor. The initial conditions
taken close enough to the unstable fixed point instead of providing inflationary de Sitter like solutions give power law
trajectories which never collapse into the stable vacuum state
intersecting the $\ddot{\phi}=\infty$ curve instead.

In other cases the behavior is similar to the previous case with $\beta=m^2_{Pl}$. Again no inflation is observed, 
and the slow rolling parameter $\varepsilon$ is
always greater than one. On the other side the phase space obtained with $\gamma=10^{-1}m_{Pl}$ provides the
physical condition ensuring inflation and it remains similar to the one shown in Fig.(\ref{plot1}). All
possible cosmological behaviors of the model are given in Tab.(\ref{table}), also for the values of 
parameters not considered in the plots. 

\begin{table}[htbp]
\begin{center}
\begin{tabular}{|c|c|c|c|c|c|c|c|}\hline
 $m_{Pl}=10$ & $\beta\rightarrow{10m^2_{Pl}}$ & $m^2_{Pl}$&$10^{-1}m^2_{Pl}$ & $10^{-2}m^2_{Pl}$ 
& $10^{-3}m^2_{Pl}$&$10^{-4}m^2_{Pl}$&$10^{-5}m^2_{Pl}$\\\hline
 $\gamma\rightarrow{m_{Pl}}/\sqrt{3}$& $-$ & $+$& $+$ & $+$& $+$& $-$& $-$\\ \hline
 $m_{Pl}10^{-1}/\sqrt{3}$ & $-$ & $+$ & $+$ & $+$& $+$& $+$& $-$\\ \hline
 $m_{Pl}10^{-2}/\sqrt{3}$ & $-$ & $+$ & $-$ & $-$& $-$& $-$& $-$\\ \hline
 $m_{Pl}10^{-3}/\sqrt{3}$ & $-$ & $-$ & $-$ & $-$& $-$& $-$& $-$\\ \hline
 $m_{Pl}10^{-4}/\sqrt{3}$ & $-$ & $-$ & $-$ & $-$& $-$& $-$& $-$\\ \hline
 $m_{Pl}10^{-5}/\sqrt{3}$ & $-$ & $-$ & $-$ & $-$& $-$& $-$& $-$\\ \hline
 $m_{Pl}10^{-6}/\sqrt{3}$ & $-$ & $-$ & $-$ & $-$& $-$& $-$& $-$\\ \hline
\end{tabular}
\end{center}

\caption{ \label{table} \small A summary of the capability of providing inflation for the NCBI cosmological
model. The combination of parameters allowing inflation are indicated with a sign $+$, the wrong ones whit a
$-$. It is obvious that this model can provide inflation only if the mass of the scalar field is close the
Planck mass. With small values of $\beta$ we get the strong BI regime, it appears that in this case only some
particular configuration of parameters allow inflation.}
\end{table}

We have tested also the amount of inflation provided by the NCBI model. 
Via numerical evaluation of $\phi(t)$ it was possible to calculate the winding number of trajectoriy by means of
the relation
\begin{equation}
N\equiv \,\ln\frac{a(t_{end})}{a(t_{initial})}=\int_{t_{i}}^{t_{e}}H\,dt\,.
\end{equation}

\noindent We see that the NCBI scalar model can generate the right amount of inflation (more than 60),
if proper slow rolling conditions are imposed. The expansion rate of this model is lower than the one
obtained with the corresponding Born-Infeld relaxed lagrangian (\ref{top-hat}).

\begin{figure}[htbp]
\begin{center}
\includegraphics[width=5.6cm, height=5cm]{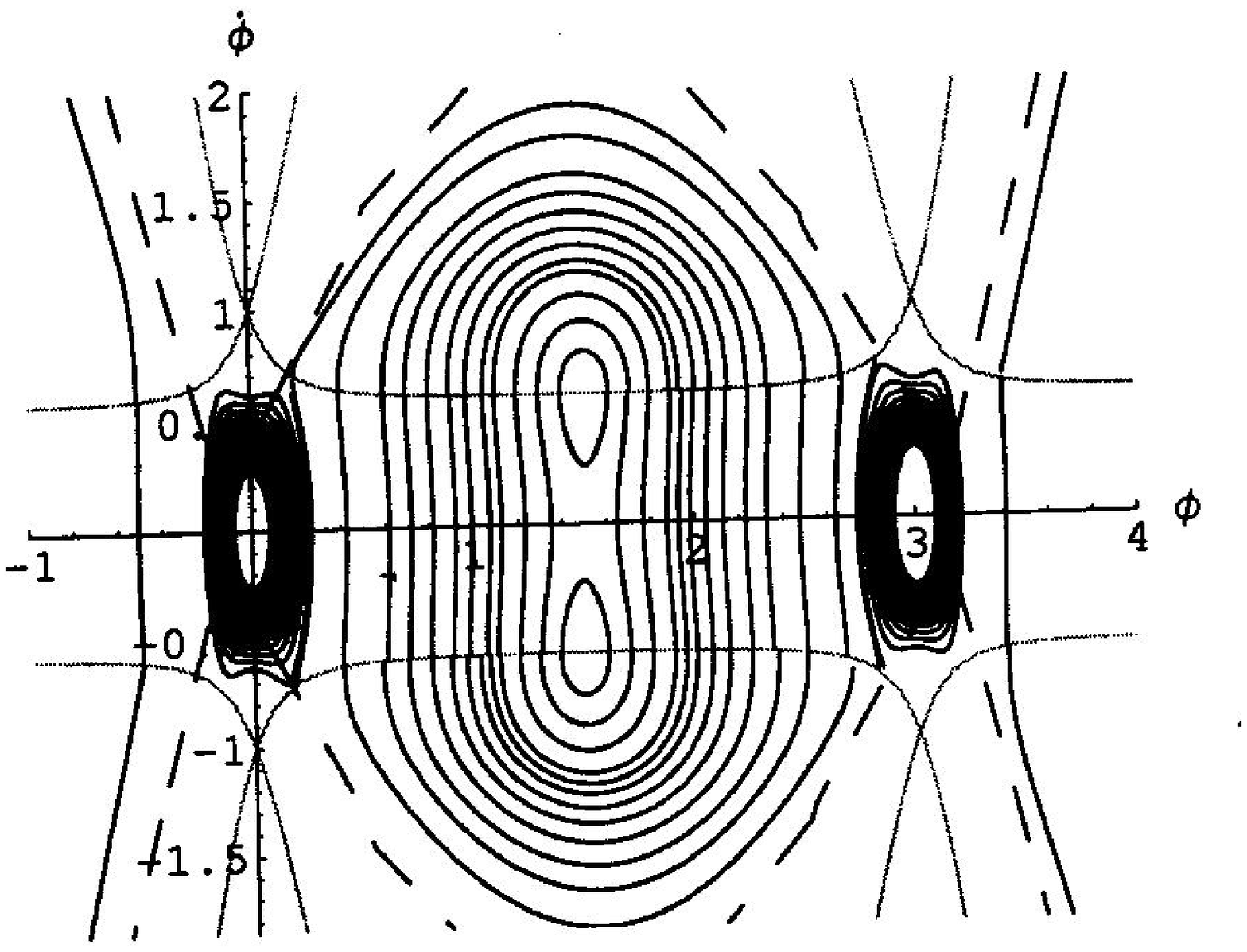}
\includegraphics[width=5.6cm,height=5cm]{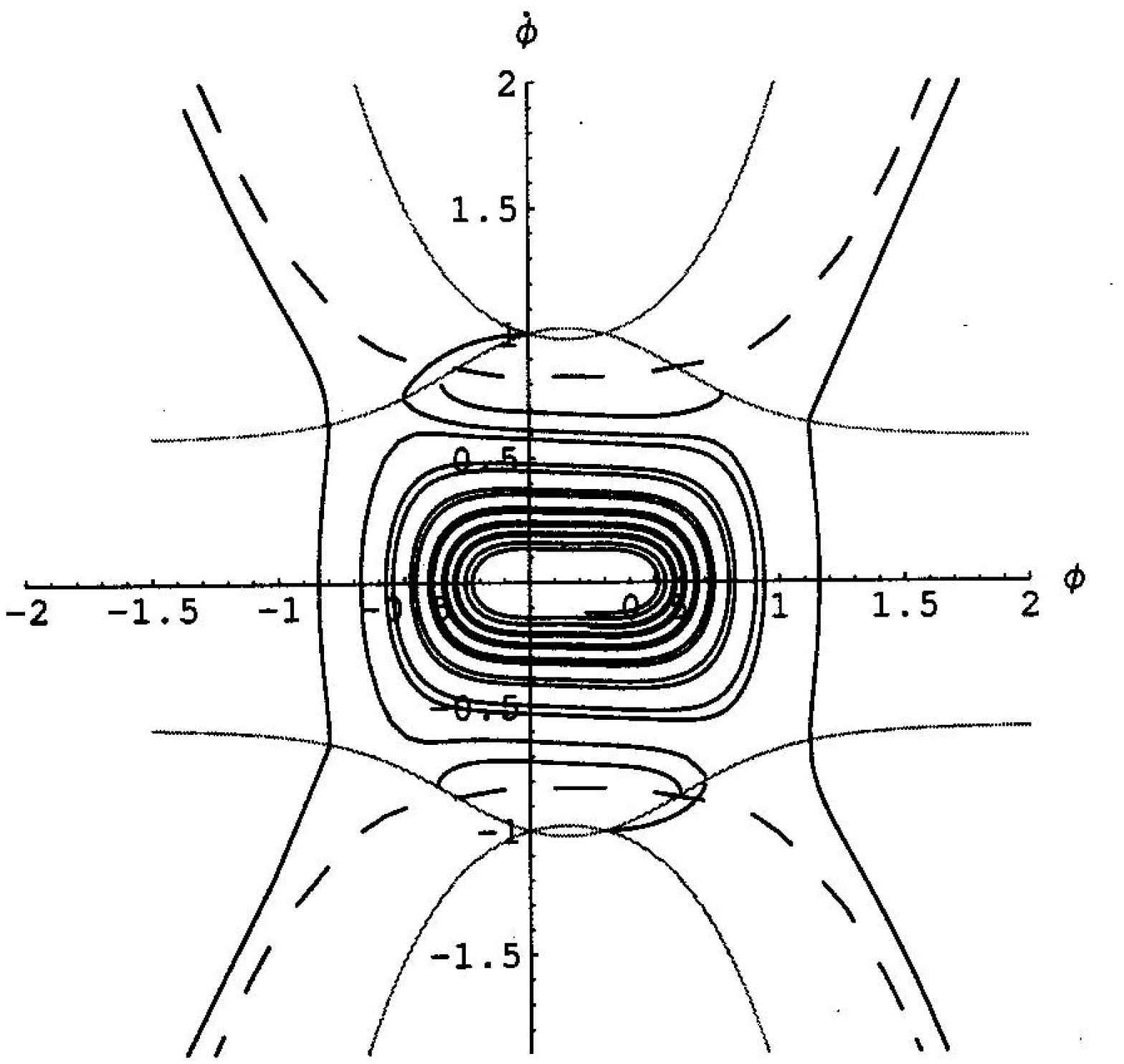}
\caption{\label{plot3-1} {\small{Phase space of the scalar field in the case $\beta=m^2_{Pl}/100$ with
$\gamma=10^{-1}m_{Pl}$ and $\gamma=10^{-2}m_{Pl}$. In the second plot the region in the middle of the curve
trajectories (small values of $\dot{\phi}$, left in white) has been plotted in detail in figure \ref{detail}. It
is obvious a behavior similar to the one found in \cite{serie:04} affected by the spiraling shape of the
trajectories due to the gravitational interaction.}}}
\end{center}
\end{figure}

\begin{figure}[htbp]
\begin{center}
\includegraphics[width=4.5cm,height=3.2cm]{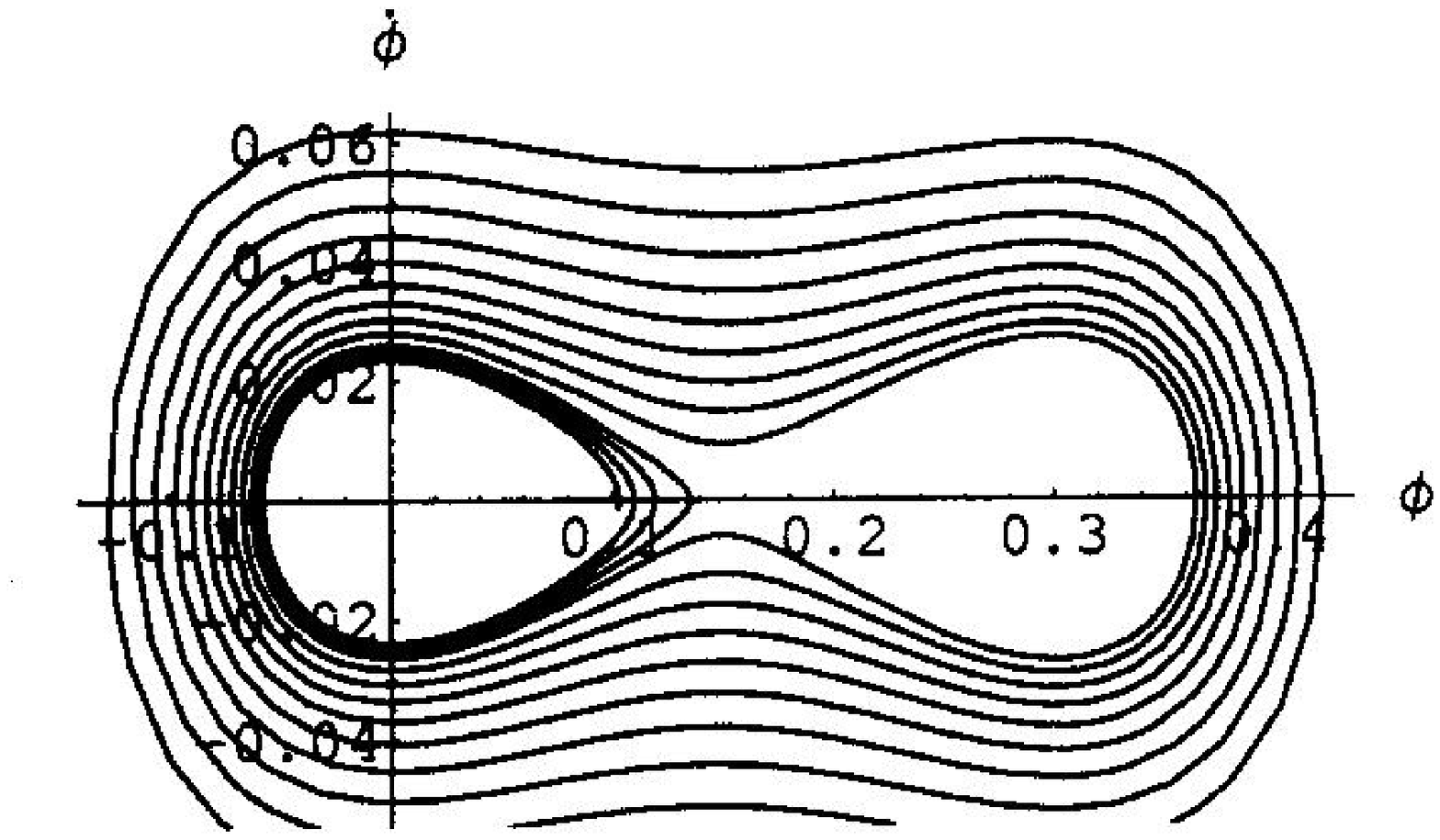}
\includegraphics[width=4.7cm,height=3.6cm]{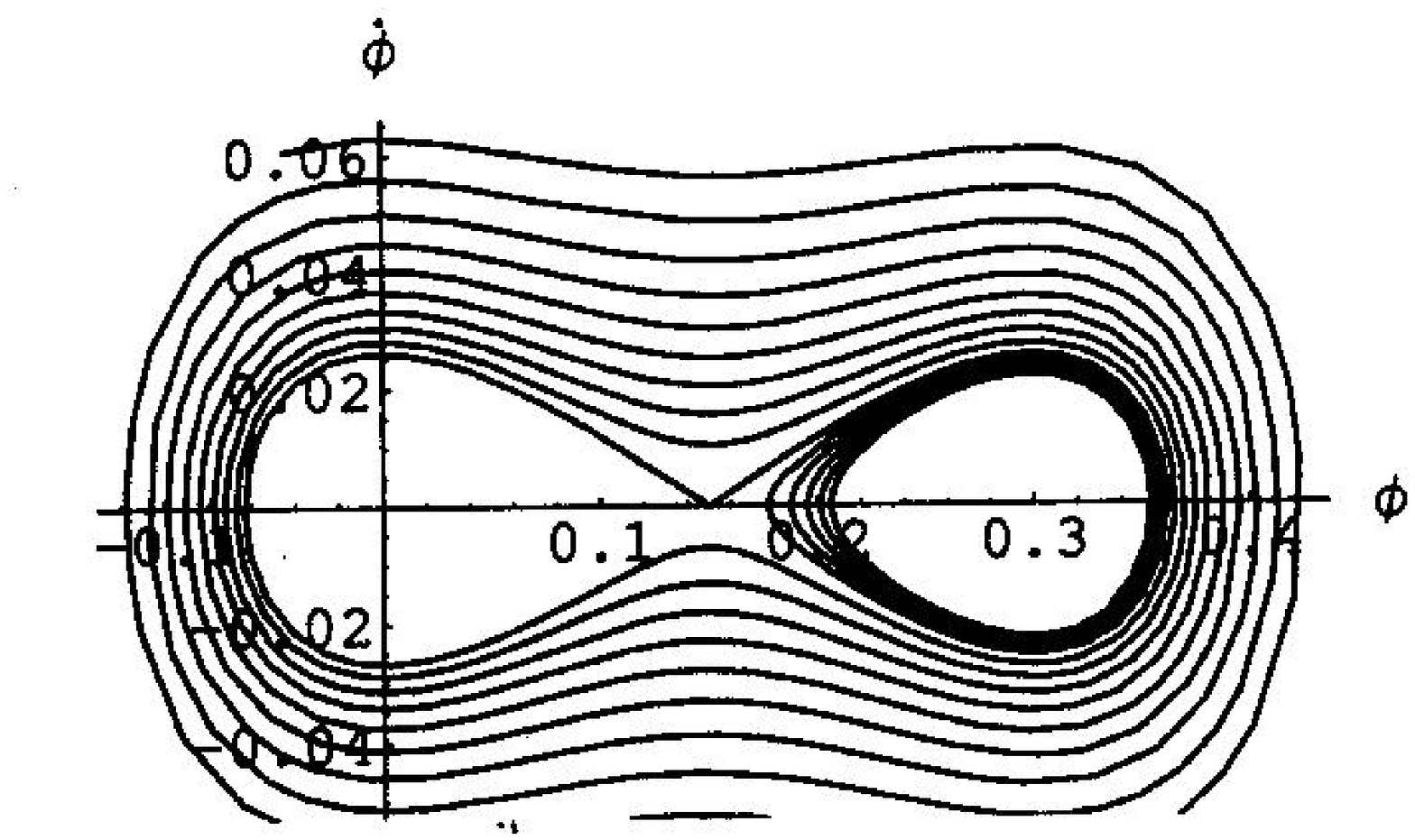}
\caption{\label{detail} {\small{The detail of the previous plot, it is showed how the wrapping of the solutions in
proximity of the critical points changes in relation to the initial conditions, left side $\phi\,=\,0,\
\dot\phi\,=\,0.16$ and right $\phi\,=\,0,\ \dot\phi\,=\,0.18$. This behavior is alternated, for
$\dot\phi\,=\,0.36$ the solutions converges to $\phi\,=\,0$ while for $\phi\,=\,0.6$ again versus
$\phi\,=\,0.33$. 
}}}
\end{center}
\end{figure}

\begin{figure}[htbp]
\begin{center}
\includegraphics[width=8.8cm,height=3.2cm]{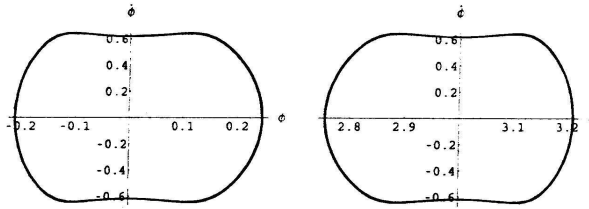}
\caption{\label{detail2} {\small{The phase portrait in the case of low gravitational coupling. A strong
similarity is found with the results of \cite{serie:04}. The gravitational interactions
break the steady state configuration of solutions; the thickness of the line is due to the very slow
spiralling of the trajectory.}}}
\end{center}
\end{figure}

Case: $\beta=10^{-2}m^2_{Pl}$.\\
This case represents the strong Born-Infeld limit in relation to the small value of the Born-Infeld parameter.
The shape of the phase space is similar to the ones proposed before, with the big feature that the allowed
region is strongly constrained. From the cosmological point of view the behavior is quite the same of the case
with $\beta=10^{-1}m^2_{Pl}$. In relation to the variation of $\gamma$ is possible to obtain inflation only for
some condition of the mass parameter. In Fig.\ref{plot3-1} we show two cases, again the solution
$\gamma=10^{-1}m_{Pl}$ shows a particular behavior as in precedence, the following cosmological properties are
similar to the previous case. Another interesting phenomena appears even in the second plot when the mass-shell
of the scalar field is close to the Planck scale. In such a case the wrapping of the phase lines evolves moving
around both the two static fixed points while the stationary configuration is achieved ending in one of this
points in relation to the initial condition without any particular rule, see Fig.\ref{detail}. This happens for
initial conditions which are chosen for all the three fixed points with several values of the velocity
$\dot\phi$.\\ Interesting plots are shown in Fig.\ref{detail2}. In this case a configuration of the parameters
is chosen in such a way that the gravitational coupling happens to be relaxed while the BI coupling is strong.
The net effect is that the spiralling behavior around stable fixed points in the left plot of
Fig.\ref{plot3-1} is more similar to the one proposed in \cite{serie:04}. Closed
trajectories shown in absence of gravity are no more possible due to the friction term and are transformed
into spirals. No significant differences are found with the variation of the mass parameter.

\begin{figure}[htbp]
\begin{center}
\includegraphics[width=4.8cm,height=3cm]{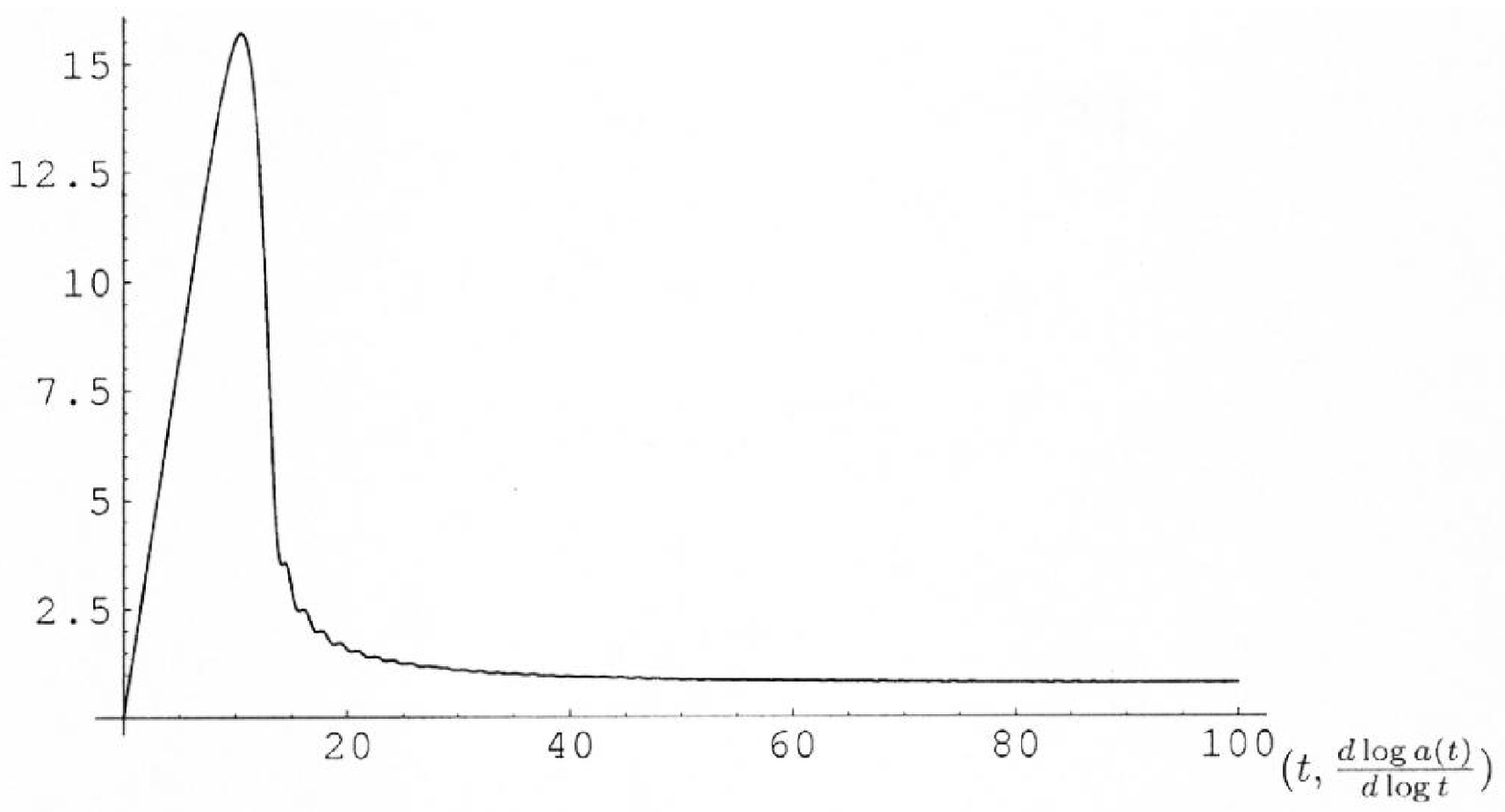}
\includegraphics[width=4.8cm,height=3cm]{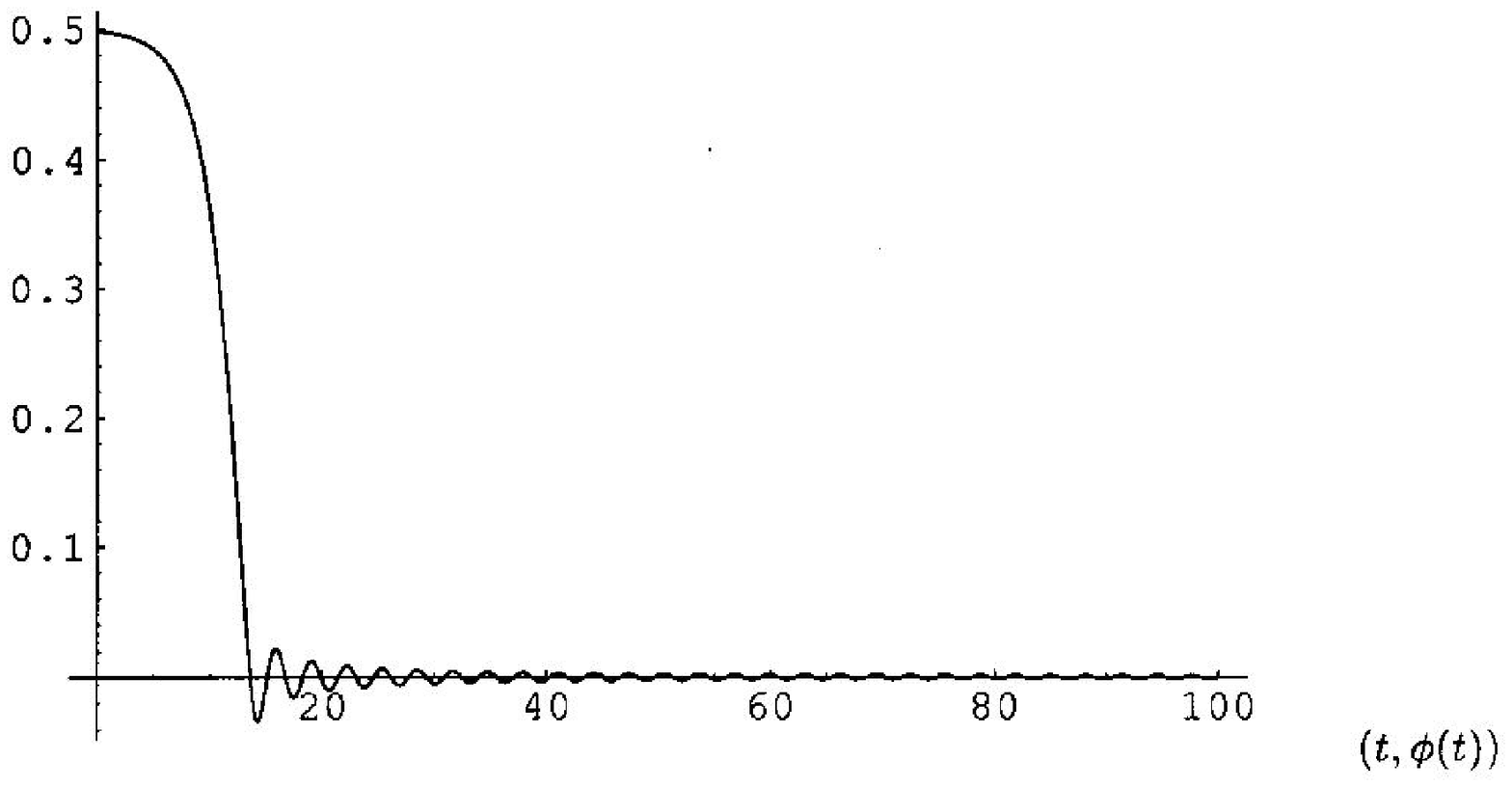}
\includegraphics[width=4.8cm,height=3cm]{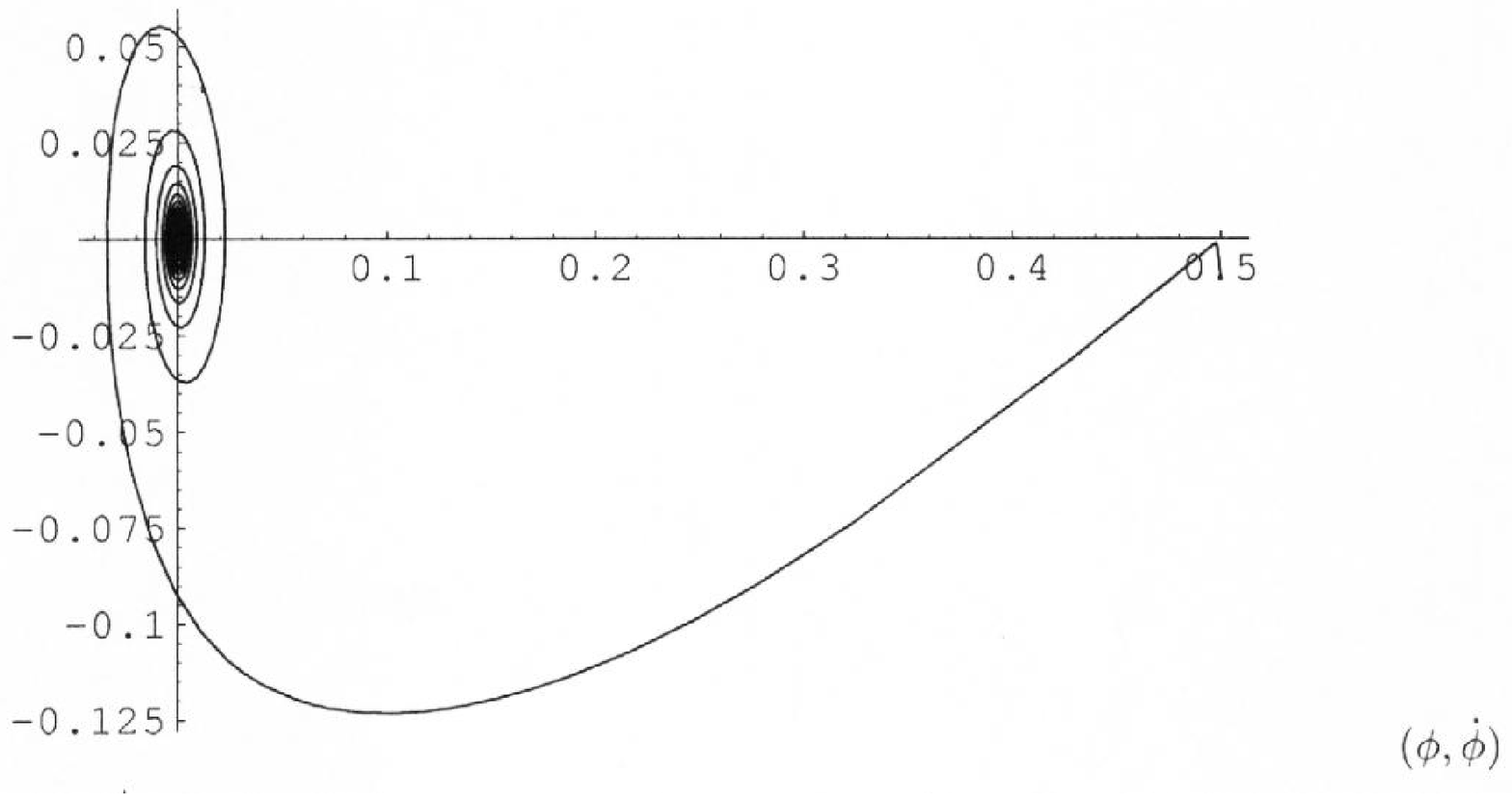}
\caption{\label{capdetail2} {\small{ The simultaneous evolution of the  Hubble parameter and of the scalar 
BI field. The last graph is the phase space plot of $\phi, \dot{\phi}$; the trajectory is spiralling towards
the singular point. }}}
\end{center}
\end{figure}

The simultaneous evolution of the cosmological scale factor $a(t)$ (through its logarithmic derivative with respect to 
$\ln t$) and of the scalar field $\phi$ is displayed in Fig. \ref{capdetail2}:

\newpage
\section{Discussion and conclusions}

In this paper we propose the cosmological analysis of a NCBI model obtained from the matrix realization of a
non-commutative geometry. This study represents the gravity coupled generalization of a model proposed in
\cite{serie:04}. To determine the cosmological properties of the model and its physical relevance in this sense
we have investigated numerically the NCBI scalar field dynamics constructing its phase space characteristic
curves. The cosmological results, in relation to several values of the parameters, are summarized in
Tab.(\ref{table}).

It is possible to observe that there are different behaviors, but in any case inflation can be recovered when
the mass of the scalar field is close to the Planck mass. A relevant difference with the $\phi^4(t)$ limit, is
that the obtained amount of inflation is smaller, although enough to solve the several shortcomings of the
standard cosmological models.

The scalar field shows a phase space which is similar to the standard case, without a BI cut-off, if $\beta$ is
big in comparison to the other parameters. In the other cases the dynamics develops inside a constrained region
around the critical points. However, the bounded regions distinctive of the Born-Infeld free theory
\cite{serie:04} are lost because of the presence of the cosmological friction term. A shape similar to the pure
BI case can be observed in Fig. \ref{detail2}, but it is obvious a slow spiralling rate of the trajectories as a
slowly decaying in the stable vacuum minima.

The most important difference with the standard case is that in the BI realm all the dynamics is
always constrained inside a well defined region. This is a persisting feature which can be found in all phase
space plots. Since for a strong BI coupling the physically significant region is very reduced  both in the
velocity and in the scalar field values, this means that such a framework  possesses an intrinsic 
condition that the scalar field should move starting only near the critical points. This is more obvious 
when the Born-Infeld parameter is small. All other initial conditions will provide trajectories which will
intersect the singular line in a finite time leading to physically meaningless solutions. In some sense this
approach contains in itself a particular choice of initial conditions without any fine tuning
procedure. 


In order to decide whether the scalar NCBI model is a viable approach of
inflation, still a large amount of work is needed, above all with respect to the observational predictions of the
model. It should be interesting to verify whether such a model can represent a good candidate for the
quintessential inflation \cite{dimoupoulos}. It is interesting to ask if the high non-linearity of the
NCBI approach can preserve safe contributes even in the late time epoch able to generate a driving force
for the new accelerating phase of cosmic expansion. These questions remain beyond the scope of the present paper.

\vskip 0.3cm
\indent
\hskip 0.5cm
{\bf Acknowledgment}
\vskip 0.3cm
\indent
A. T. is very grateful for the hospitality at the {\it Laboratoires de Physique Theorique des Liquides} of the
University Paris VI {\it Pierre et Marie Curie} where this work has been substantially developed.


\end{document}